\documentclass[12pt]{article}

\usepackage[english]{babel}
\usepackage{amsmath}
\usepackage{amsthm}
\usepackage{amssymb}
\usepackage{enumerate}
\usepackage{xspace}
\usepackage{euscript}
\usepackage{graphicx}    
\usepackage{amscd}
\usepackage{epsfig}
\usepackage{tabularx}
 \usepackage{color}
\usepackage[all]{xy}

\begin{document}
\title{\bf The scattering of a skyrmion configuration on asymmetric holes or barriers in a model Landau-Lifshitz equation}
\author{J.C. Collins\footnote{J.C.Collins@durham.ac.uk} and  W.J. Zakrzewski\footnote{W.J.Zakrzewski@durham.ac.uk} 
\\ Department of Mathematical Sciences, University of Durham, \\ Durham DH1 3LE, UK}
\date{}
\maketitle

\begin{abstract}
This current work is an extension to work previously done by the authors. The dynamics of a baby skyrmion configuration, in a model Landau-Lifshitz equation, was studied in the 
presence of  various potential obstructions. The baby skyrmion configuration was constructed from two $Q=1$ hedgehog solutions to the new baby Skyrme model in $(2+1)$ dimensions.
 The potential obstructions were created by introducing a new term into the Lagrangian which resulted in a localised inhomogeneity in the potential term's coefficient. In the
 barrier system the normal circular path was deformed as the skyrmions traversed the barrier, after which the skyrmions orbited the boundary of the system. For critical values 
of the barrier height and width the skyrmions were no longer bound although the unbound behaviour is not clearly distinct from the bound. In the case of a potential hole the 
dynamics of  baby skyrmions is dependent upon the binding energy of the system. Depending upon its value, the skyrmions' behaviour varies. The angular momentum must be modified 
to ensure overall conservation. We show that there exists a link between the oscillation in the skyrmion's
energy density and the periods of non-conservation of the angular momentum in Landau-Lifshitz models.
\end{abstract}


\renewcommand{\thesection}{\Roman{section}}

\section{Introduction}\label{intro}

There has been recent and considerable interest in the scattering of topological solitons off potential obstructions in a variety
of different models. Al-Awali et al \cite{AZ09}, \cite{AZ08}, \cite{AZ07} have considered the scattering of various topological and non-topological solitons off potential holes and barriers. Speight \cite{Spe06} has recently investigated the scattering of a topological soliton 
in a anti-ferromagnetic system, where the continuum dynamical equation resembles a second order relativistic wave equation. Both of these system have 
shown that the scattering of topological soliton off holes or barriers exhibit some interesting results. Topological solitons are, of course, classical objects. However, they describe extended 
objects and, as shown in \cite{BPZ05}, some of their properties resemble those of quantum systems. 
We have recently investigated the scattering properties of a baby skyrmion configuration in a magnetic system \cite{CZ09}, whose motion is then 
governed by the Landau-Lifshitz equation. This system showed some interesting and non-intuitive results. The potential obstructions
were placed symmetrically about the $x$-axis for all values of $x$. In this current study we shall investigate the same system with a different obstruction geometry. 

The baby skyrmions we are interested in are those of the $(2+1)$ new baby Skyrme model. The Lagrangian for the $(2+1)$ new baby Skyrme model is given by:
\begin{eqnarray}
\mathcal{L} = \frac{1}{2}\gamma_{1}\partial_{\mu}{\underline{\phi}}\cdot \partial^{\mu}\underline{\phi}
- \frac{1}{4}\gamma_{2}[({\partial_{\mu}{\underline{\phi}}\cdot\partial^{\mu}\underline{\phi}})^{2}
-(\partial_{\mu}\underline{\phi}\cdot\partial_{\nu}\underline{\phi})(\partial^{\mu}\underline{\phi}\cdot\partial^{\nu}\underline{\phi})]
- \frac{1}{2}\gamma_{3}(1-\phi_{3}^{2}) , \label{L_baby} 
\end{eqnarray}
where $\gamma_{1,2,3}$ are positive constants and $\underline{\phi}$ is a triplet of scalar fields: $\underline{\phi}(\underline{x},t) = \big\{ \underline{\phi_{i}}(\underline{x},t);i=1,2,3 \big\}$. 
The fields of this model are subject to the constraint $\underline{\phi} \cdot \underline{\phi}=1$. 
The latter two terms have been introduced to avoid the consequences of Derrick's theorem \cite{D64} and to stabilise topological soliton solutions in two dimensions. 
The condition that $\underline\phi^{2}=1$ is imposed so that the target space is the 2-sphere, such that $\underline{\phi}$ is now a map $\underline{\phi}: \mathbb {R}^{2} \rightarrow {S}^{2}$. 
For finite energy solutions it is necessary for the fields to tend to a vacuum  at infinity, where $\phi_{3}=1$ at $\infty$. This results in a compactification of $ \mathbb{R}^{2}$ so 
that $\underline{\phi}$ now takes values in the extended plane $ \mathbb{R}^{2}\cup \infty $, which is topologically equivalent to  $S^{2}$. The constraint equation $\underline\phi^{2}=1$ 
and the boundary condition at infinity results in the field $\underline \phi$ becoming a non-trivial map $\underline{\phi}:S^{2}\rightarrow S^{2}$. Each soliton solution is grouped into a 
different homotopy class according to the winding number, or topological charge, of this map. The topological charge Q is given by:
\begin{equation}
{Q}=\frac{1}{8 \pi}\int_{\mathbb{R}^{2}}d^{2}x\epsilon_{ij} \underline{\phi}\cdot(\partial_{j}\underline{\phi} \times \partial_{i}\underline{\phi}) , \label{top}
\end{equation}
 where the indices $i,j$ run over the space coordinates and $Q \in \mathbb{Z}$. The topological soliton solutions of the new baby Skyrme model are called baby skyrmions. Baby skyrmion solutions  
for this model are constructed using the Hedgehog anzatz and must be found numerically. This problem reduces to solving a second order differential equation for the profile function $f(r)$. The
 configuration we use in our investigations consists of two single charged, $Q=1$, baby skyrmions. The construction of these solutions was described in our previous work \cite{CZ09}. Static 
solutions of the new baby Skyrme model are static solutions of the Landau-Lifshitz model. It has been shown that the dynamics of two interacting baby skyrmion solutions in a Landau-Lifshitz model, 
resemble the behaviour of the experimentally observable magnetic bubbles \cite{PZ96}, \cite{PZ95}. The Landau-Lifshitz equation and its constraint are given by:

\begin{equation}
\frac{\partial\underline{\phi}}{\partial{t}}= \underline{\phi} \times-\frac{\delta{E}}{\delta\underline{\phi}}, \, \, \, \underline{\phi}^{2}=1 , \label{LL}
\end{equation}
 where $E$ is the energy functional written as:
\begin{equation}
E =\int_{\mathbb{R}^{2}}d^{2}x \mathcal{E} , \label{W} \nonumber
\end{equation}
and $\mathcal{E}$ is the static energy density of the new baby Skyrme model given by:
\begin{equation}
\mathcal{E} =\frac{1}{2}\gamma_{1}\partial_{i}{\underline{\phi}}\cdot \partial_{i}\underline{\phi}
+ \frac{1}{4}\gamma_{2}[({\partial_{i}{\underline{\phi}}\cdot\partial_{i}\underline{\phi}})^{2}
-(\partial_{i}\underline{\phi}\cdot\partial_{j}\underline{\phi})(\partial_{i}\underline{\phi}\cdot\partial_{j}\underline{\phi})] + \frac{1}{2}\gamma_{3}(1-\phi_{3}^{2}) . \nonumber
\end{equation}

Analysis of the dynamics in Landau-Lifshitz systems has been greatly simplified by the work of Papanicolaou and Tomaras \cite{PT91}, who constructed unambiguous conservation laws for the system governed by (\ref{LL}). In their work they found that the important quantity was the topological charge density $q$:
\begin{equation}
{q}=\epsilon_{ij} \underline{\phi}\cdot(\partial_{j}\underline{\phi} \times \partial_{i}\underline{\phi}). \label{topdens}
\end{equation}
Some of the conservation laws can be constructed as a moment of q. They involve:
\begin{eqnarray}
l&=&\frac{1}{2} \int_{\mathbb{R}^{2}}d^{2}x \underline{x}^{2}q  ,\label{l}\\
m&=&\int_{\mathbb{R}^{2}}d^{2}x (\phi_{3}-1)   \label{m}, \\
J&=&l+m ,
\end{eqnarray}
where $l$ is the orbital angular momentum, $m$ is the total magnetization in the third direction and $J$ is the total angular momentum.
 Conservation laws for the system were constructed by examining the time evolution of q:
\begin{equation}
{\dot {q}}=-\epsilon_{ij}\partial_{i}\partial_{l}\sigma_{jl} , \label{qdot}
\end{equation}
where $\partial_{l}\sigma_{jl}$ can be written in terms of the energy functional $E$:
\begin{equation}
\partial_{l}\sigma_{jl}=\left ( \frac{\delta E}{\delta \underline{ \phi}}\cdot{\partial_{j} \underline \phi}\right ) .
\end{equation}
Taking an explicit time derivative of (\ref{l}) gives:
 \begin{equation}
\dot{l}=\frac{1}{2} \int_{\mathbb{R}^{2}}d^{2}x\underline{x}^{2}\dot {q}.\label {ldot}
\end{equation}
 The guiding centre coordinate $\underline{R}$ of the soliton is defined as the first moment of the topological charge density $q$:
\begin{equation}
\underline{R}= \frac{1}{4 \pi Q} \int_{\mathbb{R}^{2}}d^{2}x \underline{x} q . \label {R}
\end{equation}

%

In our construction of the potential obstructions we adopt a similar approach to \cite{BPZ05}, \cite{CZ09} and introduce a term into the 
Lagrangian (\ref{L_baby}) which vanishes in the vacuum state $\phi_{3}=+1$. The obstructions need to be introduced in this 
manner so the tails of the skyrmions are not changed by the obstruction. Therefore, as in \cite{CZ09}, the additional potential 
term added to the new baby Skyrme Lagrangian (\ref{L_baby}), modifies the potential coefficient such that it now depends on the space coordinates. 
The introduction of this term implies that the static part of (\ref{L_baby}) can be rewritten as:
\begin{equation}
\mathcal{L} = \frac{1}{2}\gamma_{1}\partial_{i}{\underline{\phi}}\cdot \partial_{i}\underline{\phi} -
 \frac{1}{4}\gamma_{2}[({\partial_{i}{\underline{\phi}}\cdot\partial_{i}\underline{\phi}})^{2}
-(\partial_{i}\underline{\phi}\cdot\partial_{j}\underline{\phi})(\partial_{i}\underline{\phi}\cdot\partial_{j}\underline{\phi})] 
-\frac{1}{2}\gamma_{3}(x,y)(1-\phi_{3}^{2}), \label{L_baby_gam3} 
\end{equation}
where the potential term coefficient $\gamma_{3}$ is now a function of the coordinates $(x,y)$ and the static part of (\ref{L_baby}) 
is only considered as imposed by the Landau-Lifshitz equation.
In our previous work we presented the results of our investigations into the scattering properties
of a baby skyrmion configuration off various symmetric potential obstructions. We can ask ourselves, how many 
of the properties of that system were indicative of its symmetric nature? In other words, if the symmetric nature 
of the system was to disappear, will the dynamics of the baby skyrmions be greatly changed? Hence in this paper 
we shall discuss the results and analysis of the same baby skyrmion configuration used in \cite{CZ09}, 
in the presence of an asymmetric potential obstruction of width $b$ and height or depth $\Gamma$.
 The asymmetric obstruction is introduced into the system in the same manner as the symmetric obstruction
where  $\gamma_{3}(x,y)$ of (\ref{L_baby_gam3}) in this system can be written as:
\begin{equation}
\gamma_{3}(x,y)= 1 + \Gamma \Big \{  \Theta(-x + x_{0})  \big[ \Theta(y+y_{0})-\Theta(y-y_{0}) \big] \Big\} \nonumber ,
\end{equation}
where $y_{0}=b/2$ and $x_{0}=0$. The sign of the additional potential term's coefficient, $\Gamma$, determines
 whether the obstruction is a hole, $\Gamma<0$, or a barrier, $\Gamma>0$.
 
The evolution of the dynamics of the baby skyrmion configuration, interacting with the potential obstructions, has to be done using numerical methods. The details
 of the numerical procedures used in integrating the Landau-Lifshitz equation or the discretisation of the continuum problem etc are 
discussed in the appendix. In the absence of any obstructions, $\Gamma=0$, the two baby skyrmions of the configuration orbit each other along a circular trajectory. 
The baby skyrmions orbit the centre of the configuration $(0,0)$ anti-clockwise. In the rest of the paper we shall refer to baby skyrmions as simply skyrmions.

\section{Potential barrier}

In this section we shall illustrate some of the results seen in the potential barrier system for various values of the obstruction height $\Gamma$ and 
the barrier width $b$. The physically interesting part of these systems is the initial phase of the dynamics, when the skyrmions 
are alternating between moving `on' and `off' the barrier. The motion here is very similar to that seen in the symmetric system of \cite{CZ09}. 
Due to the asymmetry of the system the skyrmions 
no longer execute a symmetric path about the obstruction. Once the skyrmions have drifted away from the potential barrier they 
move along the boundary of the system. 
\begin{figure}[p]
\begin{center}
\unitlength1cm \hfil
\begin{picture}(13,13)
 \epsfxsize=10cm \put(1.75,8){\epsffile{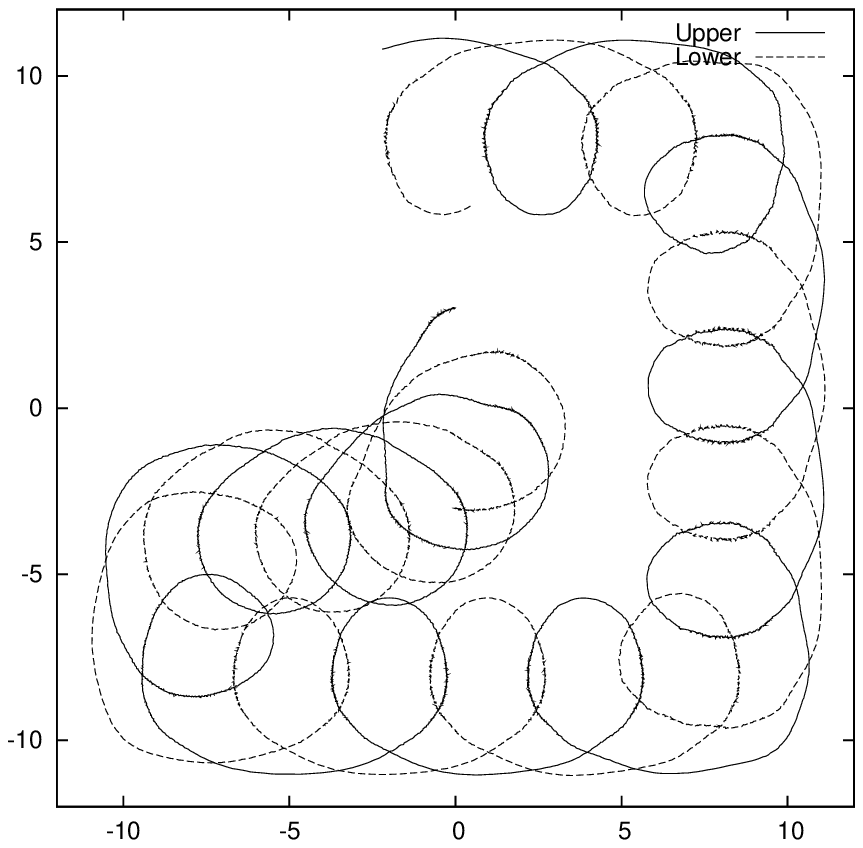}}
 \epsfxsize=10cm \put(7,0.25){\epsffile{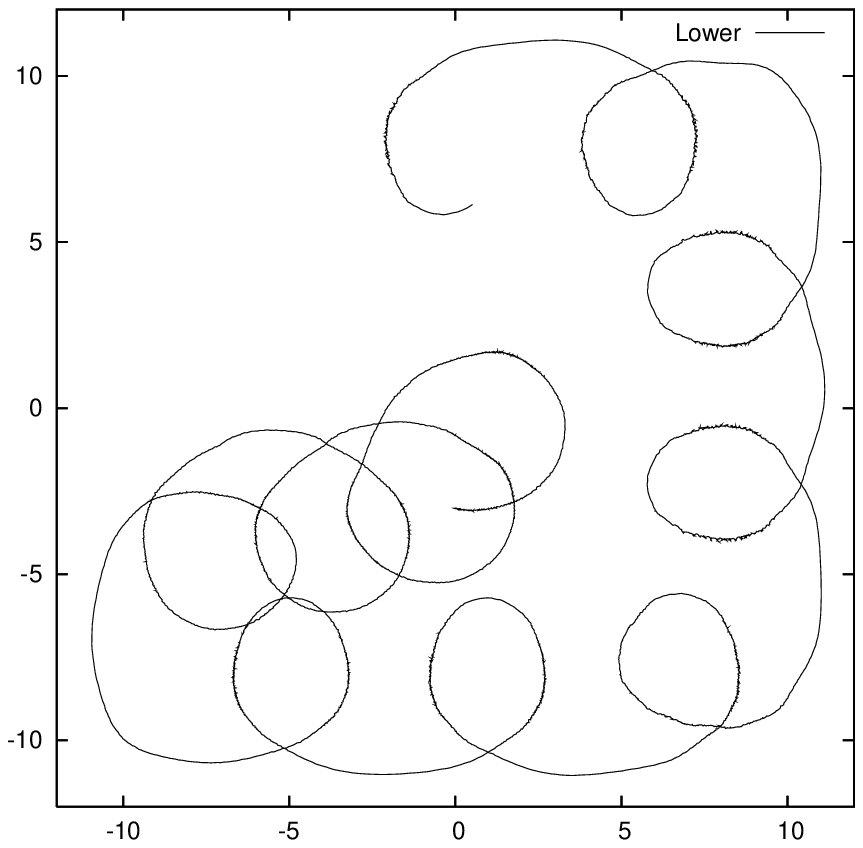}}
 \epsfxsize=10cm \put(0,0.25){\epsffile{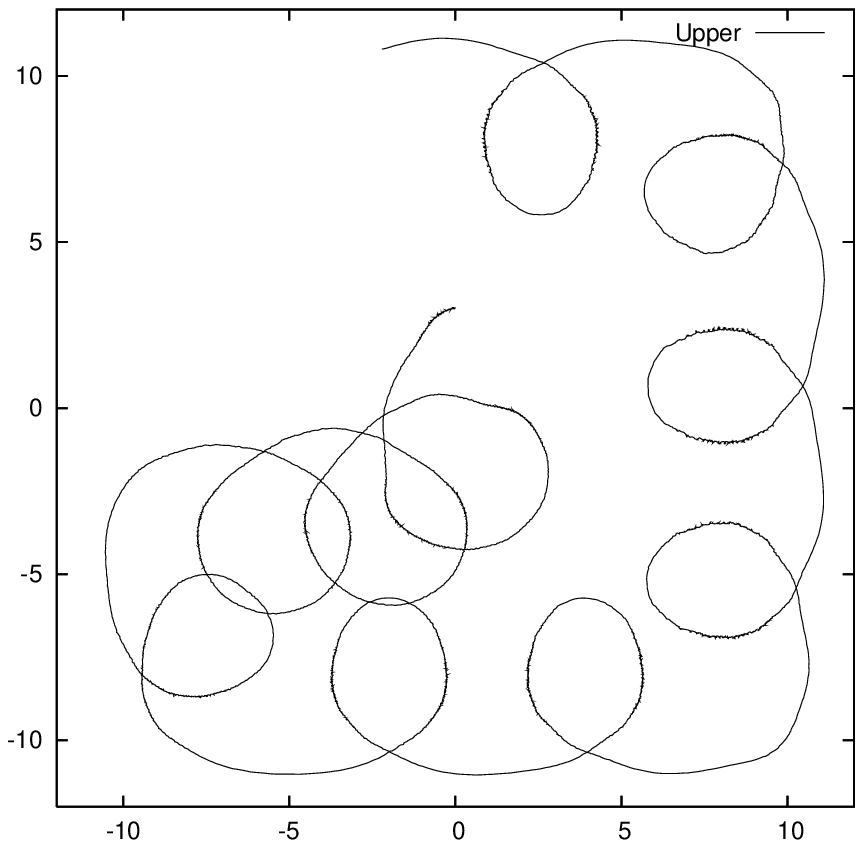}}

\put(3.75,0){a}
\put(10.75,0){b}
\put(6.75,7.5){c}
\end{picture}
\caption{\label{bar b2_025} Plots of the trajectories of the upper (a) and lower (b) skyrmions in a system with an
 asymmetric potential barrier of width $b=2$ and height $\Gamma = 0.25$, where the horizontal 
axis represents the $x$ coordinate and the vertical axis the $y$ coordinate. }
\end{center}
\end{figure}

\begin{table}
  \begin{center}
 \begin{tabular}{|c|c|c|c|}

\hline
$\Gamma$ &  $E_{2}/8\pi$  & $E_{1}/8\pi$ & $E_{B}/8\pi $ \\
\hline \hline
0.1 & 2.1298 & 1.0700 & -0.0102 \\
0.2 & 2.1319 & 1.0706 & -0.0093 \\
0.3 & 2.1340 & 1.0713 & -0.0086 \\
0.4 & 2.1361 & 1.0719 & -0.0077 \\
0.5 & 2.1382 & 1.0725 & -0.0068 \\
1.0 & 2.1487 & 1.0757 & -0.0027 \\
1.1 & 2.1508 & 1.0763 & -0.0018 \\
1.2 & 2.1529 & 1.0770 & -0.0011 \\
1.3 & 2.1550 & 1.0776 & -0.0002 \\
1.4 & 2.1571 & 1.0782 & +0.0007 \\
1.5 & 2.1592 & 1.0789 & +0.0014 \\
\hline

\end{tabular}

\caption{\label{d=6 barrier table} Table showing the variation in the total energy and binding energy for a $d=6$ two 
skyrmion configuration interacting with a potential barrier of width $b=2$ for various values of $\Gamma$ .}
\end{center}

\end{table}

The analysis of the skyrmion scattering in the barrier system can be divided into the sections where the skyrmions are far from
 the obstruction and when they are executing their circular motion `on' or near the barrier. 
The initial phase of the dynamics corresponds to the latter of these two sections.
Fig. \ref{bar b2_025} shows multiple plots of a $d=6$ skyrmion configuration interacting with an asymmetric barrier of width $b=2$ and
 depth $\Gamma=0.25$. The skyrmions centres are initially located at $(0,\pm 3)$ where we refer to the skyrmion initially placed at $(0,3)$
as the upper skyrmion and $(0,-3)$ as the lower one. 
Fig. \ref{bar b2_025}a and b show the trajectories of the upper and 
lower skyrmions over a time length of $3500$ secs. Fig. \ref{bar b2_025}c shows the two trajectories plotted together. It can be 
seen from the first two plots that initially the two skyrmions try to execute their normal circular motion around each other. 
As the skyrmions approach the obstruction, their path is deformed due to the barrier. The trajectory of the skyrmions in this 
system suggests that the skyrmions still form a bound state for these 
values of $\Gamma$ and $d$. The binding energy of a two-skyrmion configuration in the presence of a potential obstruction is 
denoted by $E_{B}$ and can be defined by the following: if the energy of the skyrmion configuration in the presence of a barrier is 
denoted by $E_{2}$ and $E_{1}$ is the energy of a single skyrmion placed at the same position as one of the skyrmions in $E_{2}$, then the binding, or the interaction energy, 
is given by:
\begin{equation}
 E_{B}=E_{2}-2E_{1} .\label{EB}
\end{equation} 
Table \ref{d=6 barrier table} shows the variation in the binding energies for various values of $\Gamma$ for a $d=6$ skyrmion 
configuration, interacting with a potential barrier of width $b=2$. Here for $\Gamma=0.25$ the binding energy of the skyrmions $E_{B}<0$ 
and thus the skyrmions still form a bound state, which the plots in Fig. \ref{bar b2_025} confirm.

\begin{figure}[h]
\begin{center}
\unitlength1cm \hfil
\begin{picture}(10,10)

\epsfxsize=14cm \put(-3,0){\epsffile{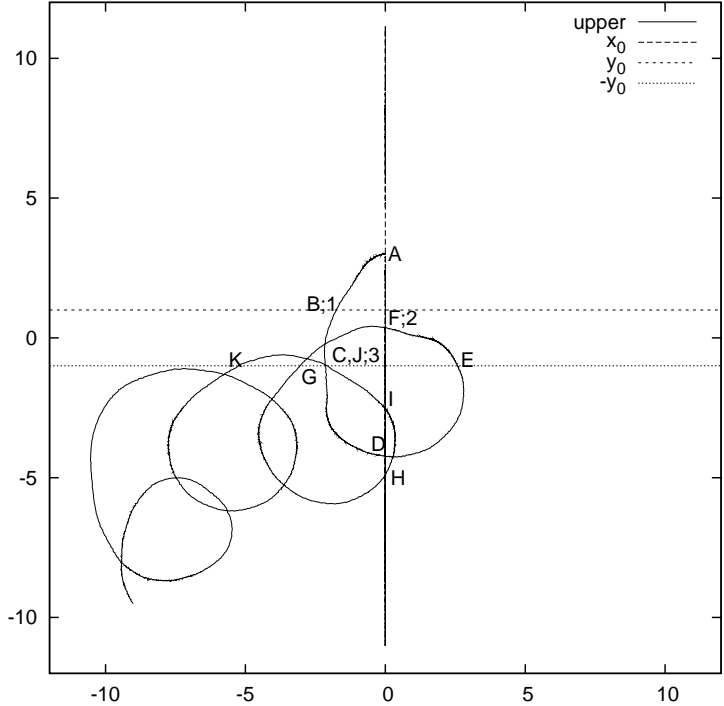}}

\end{picture}
\caption{\label{upper label} Plot of the trajectory for the upper skyrmion of a $d=6$ configuration interacting with an asymmetric potential barrier of width $b=2$ and $\Gamma = 0.25$, where the horizontal 
axis represents the $x$ coordinate and the vertical axis the $y$ coordinate.}
\end{center}
\end{figure}

\begin{figure}[p]
\begin{center}
\unitlength1cm \hfil
\begin{picture}(13,13)

\epsfxsize=10cm \put(0,8.25){\epsffile{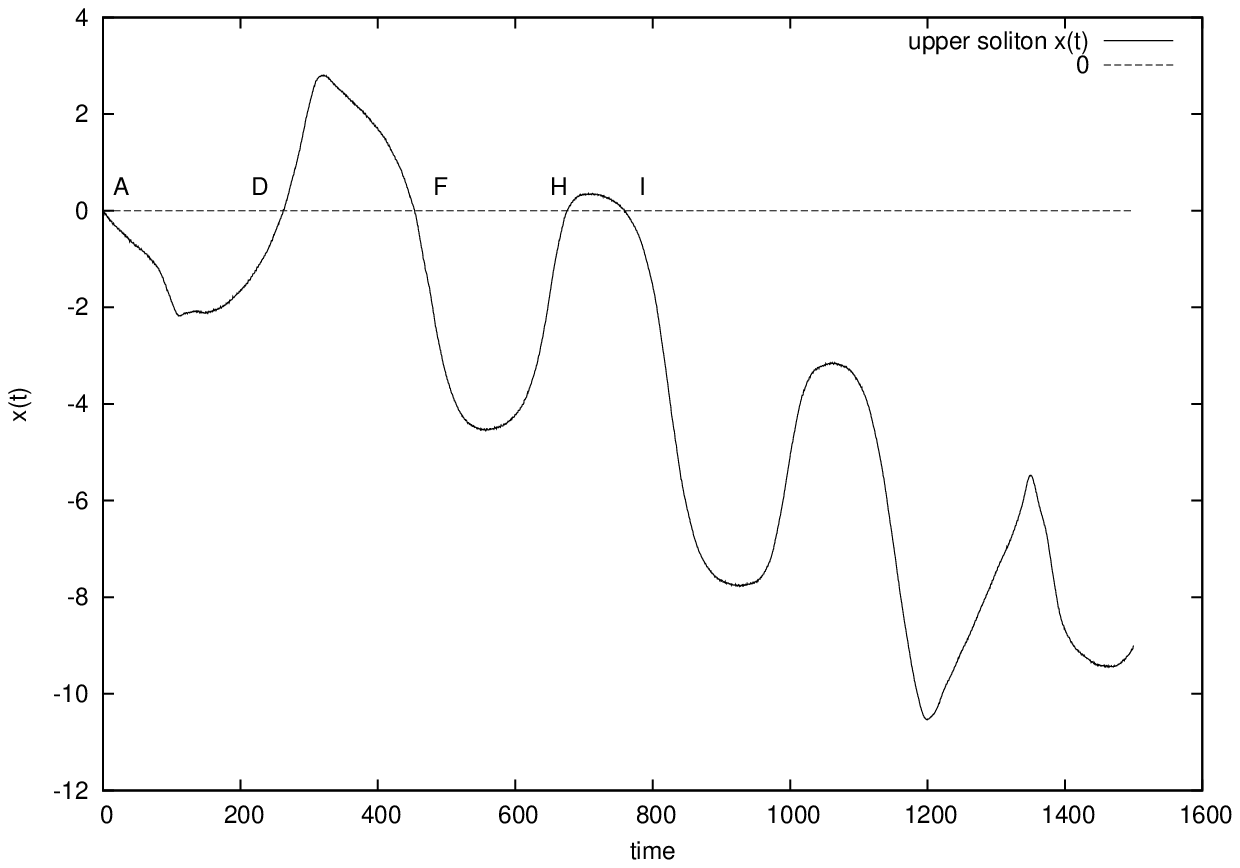}}
\epsfxsize=10cm \put(0,0.45){\epsffile{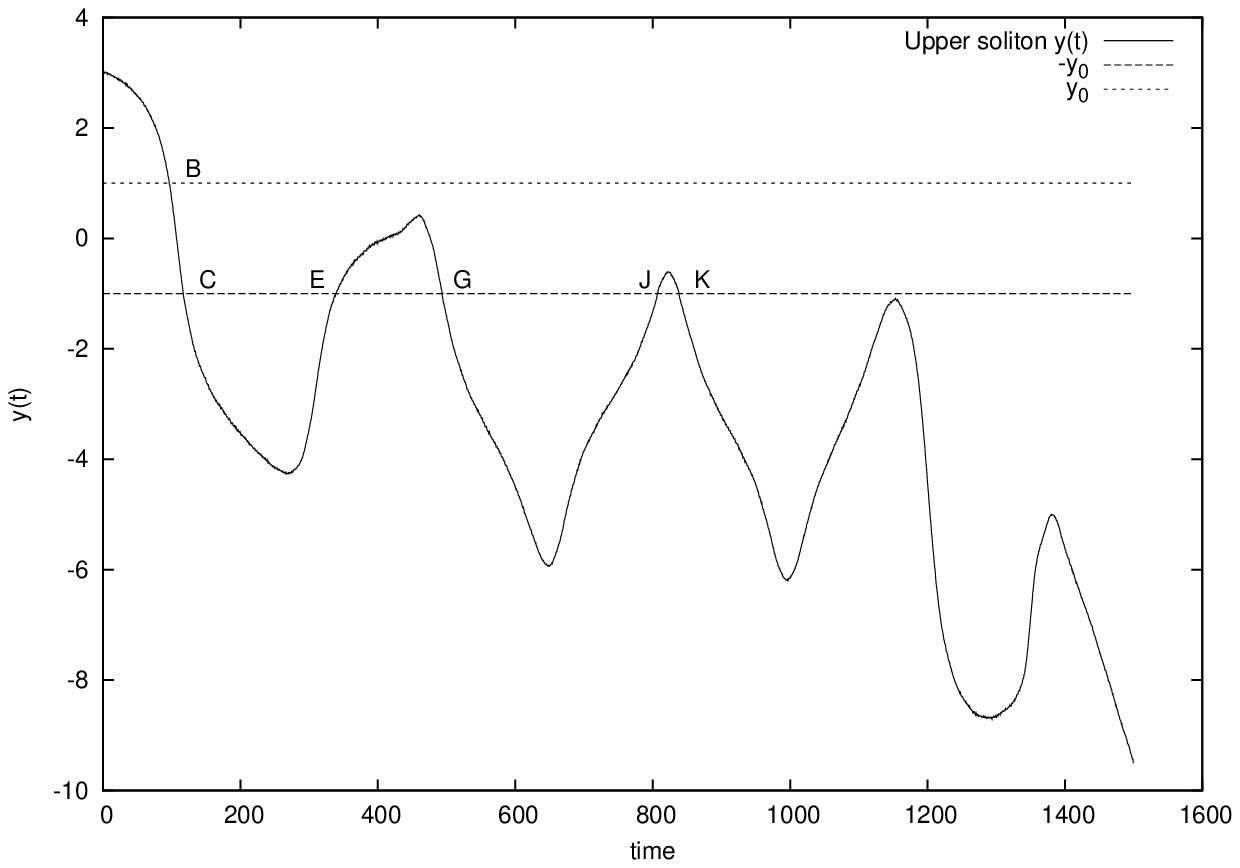}}

\put(5.25,7.75){a}
\put(5.25,0){b}

\end{picture}
\caption{\label{upper x(t)} Plot $x(t)$ and $y(t)$ for the upper skyrmion in a system with an asymmetric potential barrier of width $b=2$ and $\Gamma = 0.25$. }
\end{center}
\end{figure}

  Fig. \ref{upper label} shows the plot of the initial phase of the above system, up to $1500$ secs, with a set of labels A-K and 1-3. 
The labels A-K are reference points used to examine the motion and speed of the 
skyrmions during this initial phase. The labels 1-3 refer to the different times the skyrmions traverse the barrier. The potential
 obstruction's position is outlined by the dotted lines appearing on the plot of  Fig. \ref{upper label}. Although the 
barrier has been defined in section \ref{intro}, the lines have been continued for $x>0$ and for $y > y_{0} \, , \, y<-y_{0}$ for
 purely illustrative purposes. In order to identify the times at which the skyrmions have entered the barrier region we have 
further plotted $x(t)$ and $y(t)$ for the upper skyrmion shown in Fig. \ref{upper x(t)}. These plots are labelled 
with A-K accordingly. 

The initial path of the upper skyrmion as it approaches
 the barrier is A-B. The distance of A-B and the distance of the path over the barrier B-C are comparable. However, 
the time taken for the skyrmion to travel A-B is $\sim$ 100 secs, whereas to traverse B-C it is $\sim$ 30 secs. If we 
examine the gradient of $y(t)$ for the upper skyrmion, we can see that the gradient rapidly increases and then decreases as 
the skyrmions traverse the barrier. The speeding up of the skyrmions as they traverse a potential barrier was seen in 
the analysis of systems with a symmetric potential obstruction. In the further two scattering paths the times taken to 
traverse the barrier is $\sim$ 30 secs for F-G, and $\sim$ 30 secs for J-K. It can be shown that the lower skyrmion traverses the barrier in similar 
times as the upper skyrmion, for paths 1, 2 and 3.


%
%

\begin{figure}[!h]
\begin{center}
\unitlength1cm \hfil
\begin{picture}(8,8)

\epsfxsize=10cm \put(-2,0.25){\epsffile{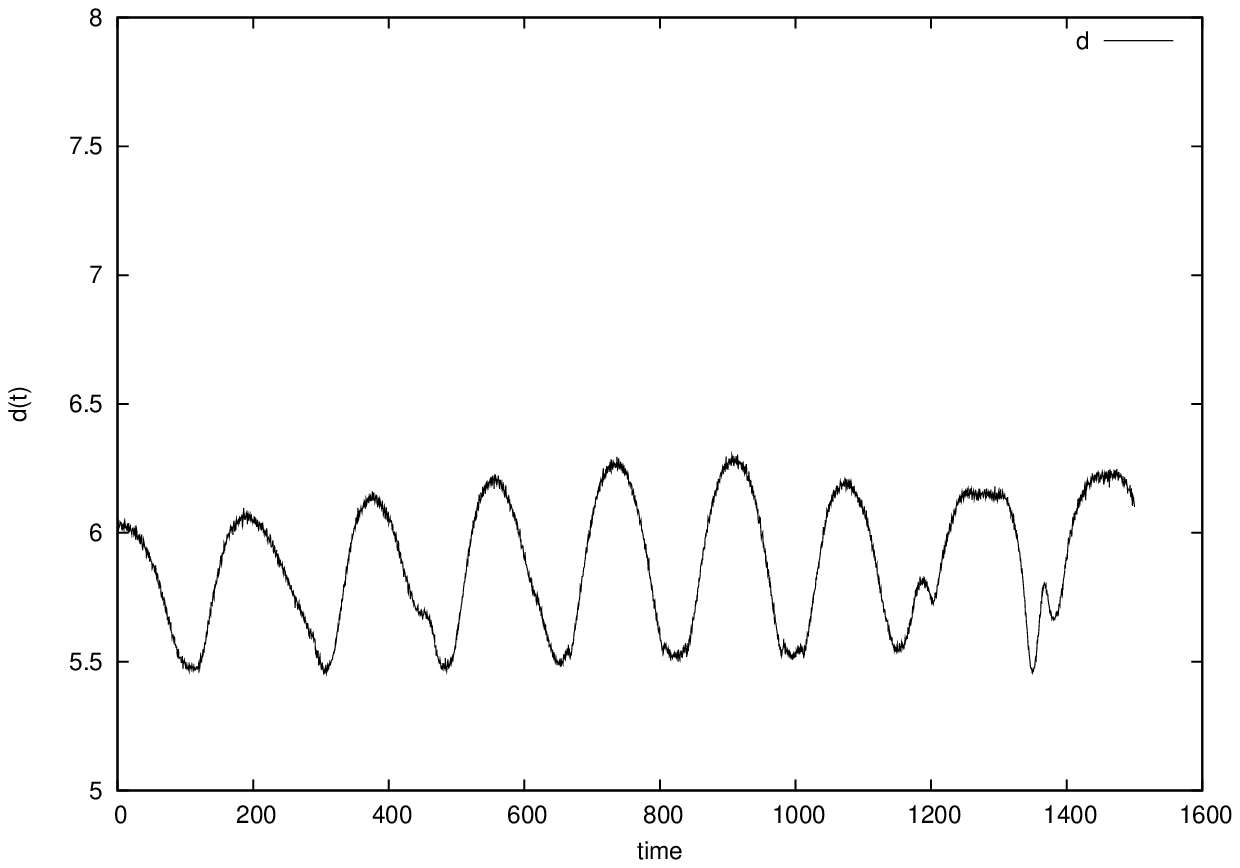}}

\end{picture}
\caption{\label{b2 d(t)}Plots of the distance of separation, $d$, between two skyrmions in a system with an asymmetric potential 
barrier of width $b=2$ and $\Gamma = 0.25$.  }
\end{center}
\end{figure}

Fig. \ref{b2 d(t)} shows a plot of the skyrmion separation as a function of time, $d(t)$, for the first 1500 secs, 
with an initial separation of $d=6$. As the skyrmions move `on' and `off' the barrier, the distance of 
separation decreases and increases accordingly. The minima of $d(t)$ correspond to the times when one of the skyrmions 
is on the barrier. The skyrmions, initially placed in the attractive channel, compensate for the increase in potential
 energy by reducing their distance of separation. Once off the barrier the skyrmion configuration returns to its initial 
state and the skyrmions separate once again. This process repeats with each
of the different paths 1-3.


\subsection{Away from the barrier}\label{boundary}

\begin{figure}[!htpb]
\begin{center}
\unitlength1cm \hfil
\begin{picture}(10,10)

\epsfxsize=14cm \put(-3.5,0.25){\epsffile{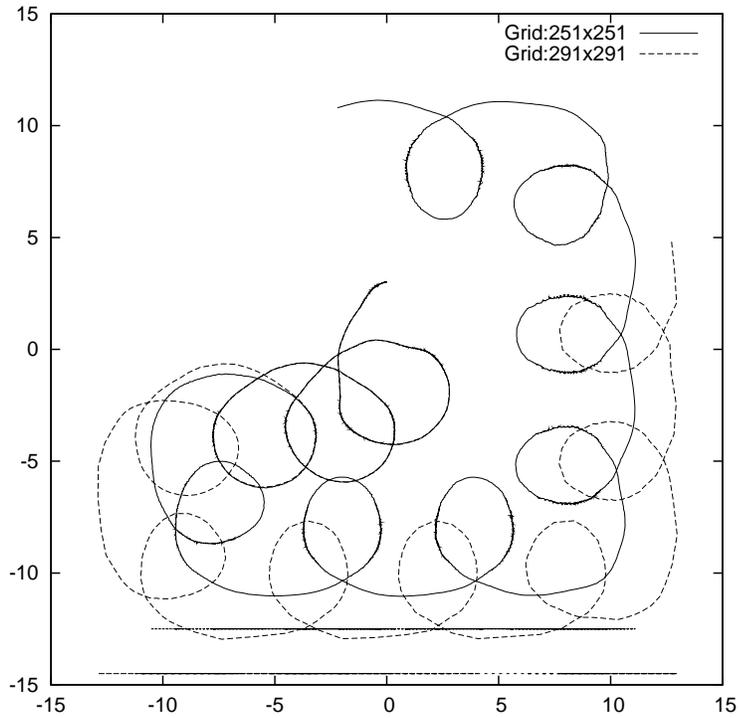}}

\end{picture}
\caption{\label{upper grid} Plot of the trajectories of the upper skyrmions, where both skyrmions are initially at $(0,3)$, in a system with an asymmetric potential barrier
 of width $b=2$ and $\Gamma = 0.25$ for grid sizes 251 $\times$251 
 and  291 $\times$ 291 with the boundary of different grid size shown. The horizontal 
axis represents the $x$-coordinate and the vertical axis the $y$-coordinate. }
\end{center}
\end{figure}

Another interesting feature of this system is the way the skyrmions avoid the boundary and potential obstruction 
after the initial phase described previously. After the transient period where the skyrmions traverse the barrier, 
they eventually settle down to a steady state where the skyrmion configuration moves about the grid avoiding the barrier 
and the boundary. During this steady state the skyrmions continue to execute the circular motion about the configuration 
centre when they are close enough to interact. The skyrmions have a strong repulsion from the boundary and keep a certain 
distance from it. The skyrmions' motion in this phase is determined by the boundary. If we increase the size of the boundary, 
after performing a similar initial phase, the skyrmions continue to move around the edge of the boundary. This is illustrated 
in Fig. \ref{upper grid} which shows the upper skyrmion of the same system of $b=2$ and $\Gamma=0.25$, plotted with an identical upper 
skyrmion in a system with the same values of $b$ and $\Gamma$ but with a grid size of 291 $\times$ 291 points. The boundary of the two 
different grid sizes is also shown on Fig. \ref{upper grid}.  The larger grid 
means that the boundary of the system is further away than for the system with a smaller grid size. This plot
confirms that the skyrmions' trajectory away from the barrier is dependent upon the boundary. Thus the physically relevant part 
of the dynamics i.e. the phase unique to the problem of interest, is the initial phase before the skyrmions drift along the boundary's edge.


In the previous sections we described the general system of a bound skyrmion configuration interacting with a potential barrier. 
This system was chosen because it illustrated many of the generic features of the skyrmion's 
motion in these systems either in the region near the obstruction or away from it. However, there are a few additional features which exist 
for only certain values of the barrier height $\Gamma$, and distance of skyrmion separation $d$, and shall be presented here.

\begin{figure}[htbp]
\unitlength1cm \hfil

\begin{picture}(13,13)
 \epsfxsize=10cm \put(0,8){\epsffile{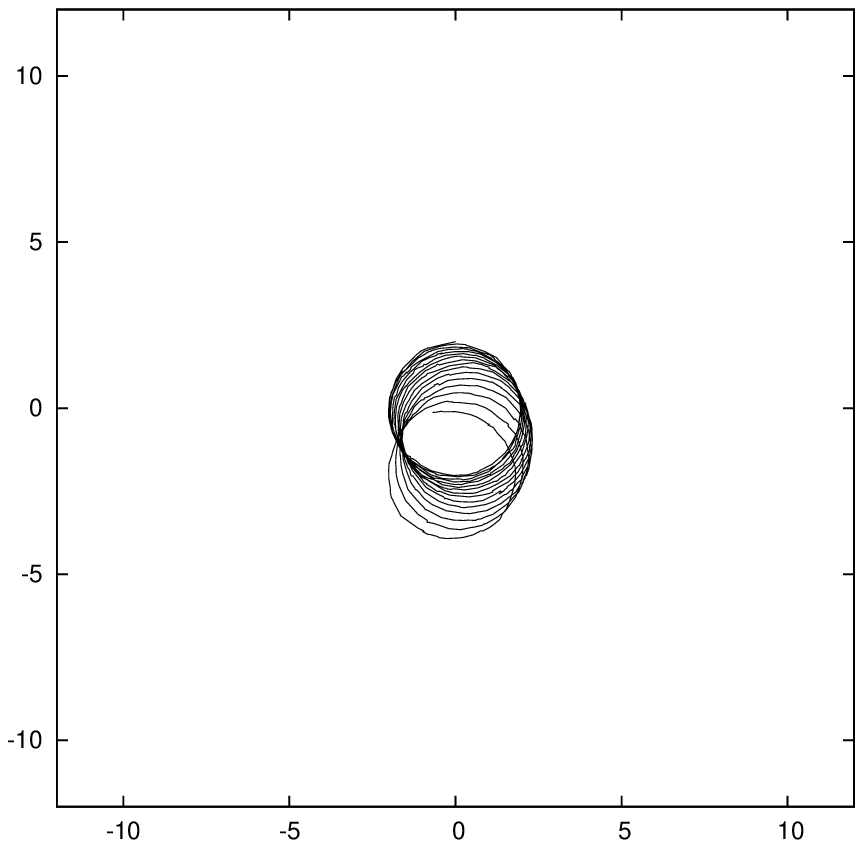}}
 \epsfxsize=10cm \put(7,8){\epsffile{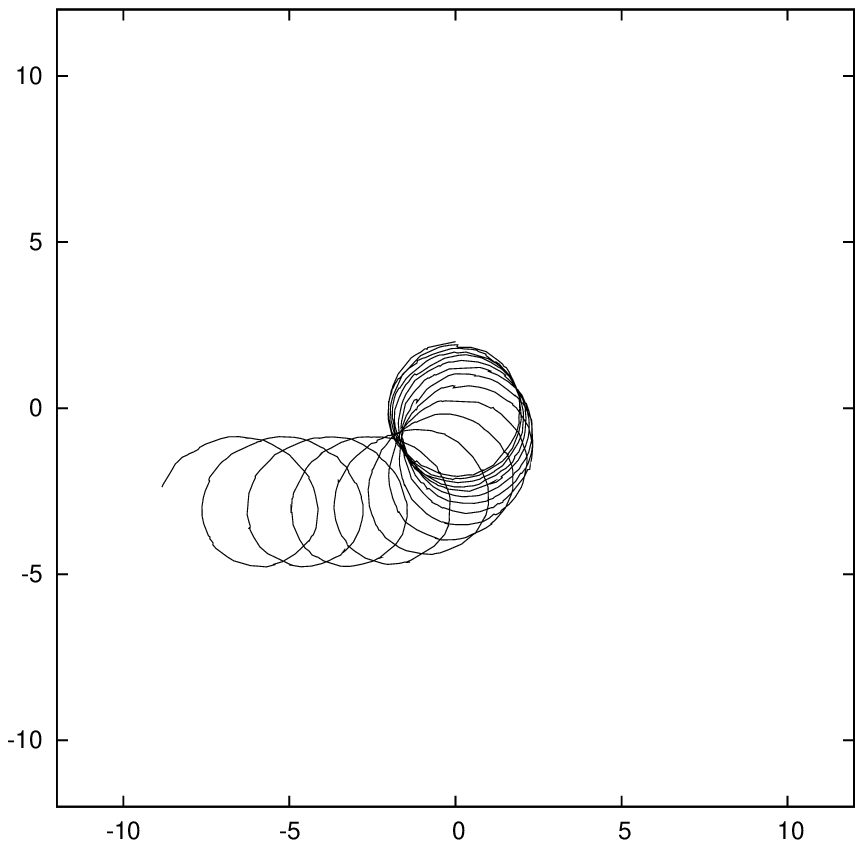}}
 \epsfxsize=10cm \put(0,0.25){\epsffile{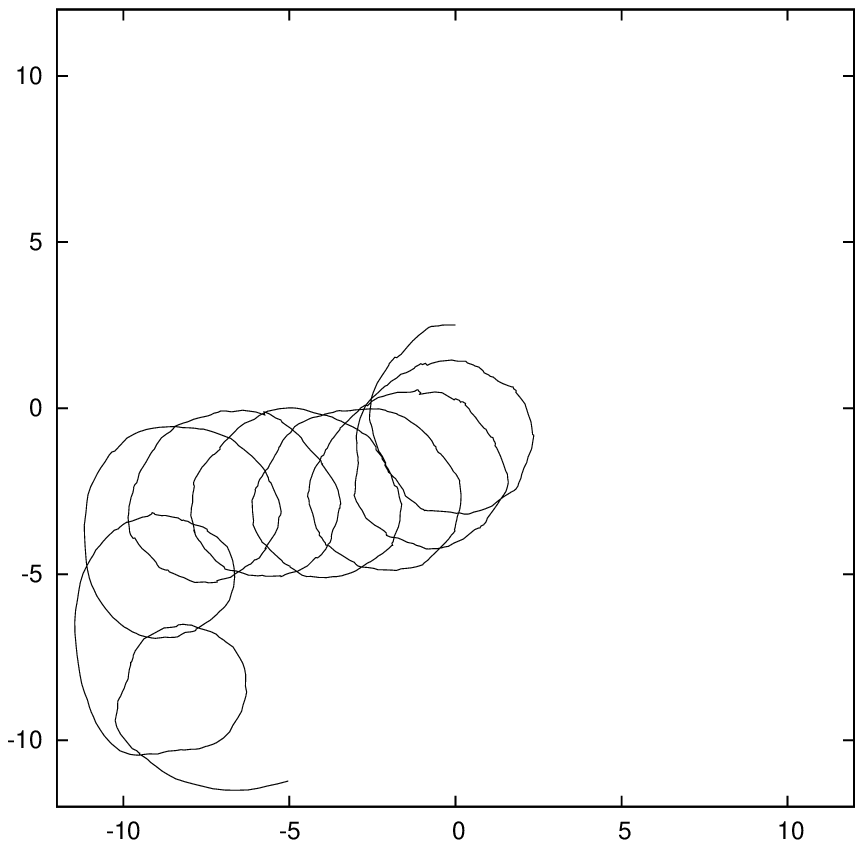}}
\epsfxsize=10cm \put(7,0.25){\epsffile{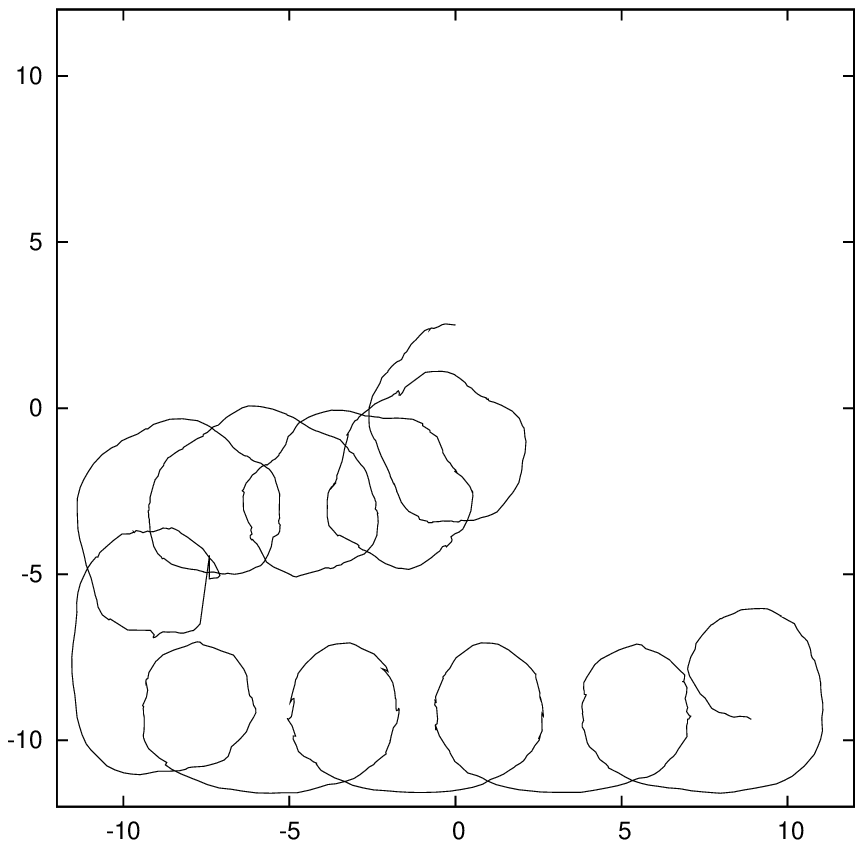}}

\put(5,0){c}
\put(12,0){d}
\put(5,7.5){a}
\put(12,7.5){b}

\end{picture}
\caption{\label{d=4 and d=5 barrier} Plots of the trajectories for the upper skyrmions 
of a system with an asymmetric potential barrier of width $b=2$ for various values of 
$\Gamma$ and $d$ : a) $\Gamma = 0.3$ and $d=4$, b) $\Gamma = 0.5$ and $d=4$, c) $\Gamma = 0.5$ and $d=5$ and  d) $\Gamma = 0.8$ and $d=5$, where the horizontal 
axis represents the $x$ coordinate and the vertical axis the $y$ coordinate. }

\end{figure}

In the $d=6$, $\Gamma=0.25$ barrier system, we observed that the skyrmions were still bound due to value of the binding energy. This 
was also reflected in the trajectories of the skyrmions in that system. The skyrmion configurations when $d<6$, due to the attractive 
channel, are more tightly bound than the $d=6$ skyrmions. In systems with small values of $\Gamma$, the skyrmions exhibit bound behaviour 
in both the binding energy and trajectories.  Away from the barrier they orbit the boundary of the system as usual.  This behaviour is 
evident in Fig. \ref{d=4 and d=5 barrier} which shows the plots of the trajectories for a skyrmion configuration with $d=4$ and $d=5$ for 
various values of the barrier height $\Gamma$. The skyrmions in these systems orbit each other much faster than the looser bound $d=6$ 
system. The trajectories of the skyrmions in the $d=4,5$ systems are slowly distorted as they execute 
their motion. As the value of $\Gamma$ increases the distortion of the skyrmion trajectory becomes more pronounced.


In the absence of any obstructions the skyrmions form a bound state. The strength of the bound state is dependent upon the separation between 
the two skyrmions. For $d=6$ the skyrmions are reasonably well 
separated and the skyrmions are not tightly bound. For large values of $\Gamma$ however, the binding energy of these skyrmion configurations can increase and the bound state 
can be broken. As the barrier height gets larger, the skyrmions are unable to overcome it. In \cite{CZ09}, in the barrier's 
transition dynamics, it was seen that the skyrmions cannot distinguish between the boundary of the system and very large barriers. 
We showed that the skyrmions scatter at an angle $\alpha$ relative to the barrier's edge. This angle $\alpha \rightarrow 0$ as $\Gamma$ 
gets larger. In the asymmetric system for a sufficiently large barrier, which the skyrmions cannot traverse, the skyrmions begin to move 
at this angle $\alpha$. Initially the upper skyrmion starts to move in the direction away from the obstruction but due to the asymmetry 
of the system it is able to interact with the lower skyrmion and sling shot around the edge of the barrier.

The trajectory of a set of skyrmions, where the bound state is broken by the obstruction, is much more difficult to define in the asymmetric system than in the symmetric 
one of \cite{CZ09}. In the symmetric system the unbound skyrmions, after being repelled from the 
barrier, were unable to move into the lower or upper semi-planes because the obstruction was placed over the whole $x$ range. In the asymmetric 
system the skyrmions are able to move into these semi-planes. Thus in the symmetric system the trajectory of an unbound configuration was clearly defined. When two skyrmions 
are close enough to interact they will perform their usual circular motion about the configuration centre. It is difficult therefore to 
distinguish between states that are unbound or those that are not. The only way one is certain the skyrmions are unbound is to observe 
states for a sufficiently large values of $\Gamma$. In those cases, after the skyrmions have moved around the the obstructions, we observe 
the skyrmions moving about the boundary of the system singly. Periodically the skyrmion configuration undergoes circular motion 
when the skyrmions are close enough to interact.
 

\section{Potential hole}

In \cite{CZ09} it was observed that the dynamics of the skyrmion configuration interacting with a potential hole was largely dependent 
upon whether the skyrmions were bound or not. When the skyrmions were unbound they were free to separate from each other. During the 
separation the skyrmions moved along the edge of the potential hole. This movement continued until the skyrmions reached the physical boundary
of the system. It is here that both skyrmions underwent reflection at right angles to the boundary, traversing the hole and then 
executing an identical trajectory along the opposite edge of the the hole. The skyrmions motion continued back to the starting points 
where the system looked the same, with just an exchange of the two skyrmions. In the system of an asymmetric obstruction we would 
expect the previously mentioned symmetric trajectories to no longer persist. As indicated in previous sections, the hole exists on 
only half of the plane and hence the dynamics in this system should differ from the symmetric system. However, we expect that the 
factor determining the form of the dynamics should be the same as in the symmetric system, namely, the binding energy, $E_{B}$, defined in (\ref{EB}). 
In the following three sections we shall present and discuss the results of a two-skyrmion configuration interacting with a potential hole of 
varying widths and depths, for a number of different skyrmion separations, $d$.
The different sections are categorised according to whether the skyrmion dynamics is bound, unbound or in between. 

\subsection{Bound dynamics}

The skyrmions are initially set up in an attractive channel, which means that for close distances they form a bound state. Thus the states 
that one would presume to exhibit bound-like behaviour are those corresponding to small distances of separation and small hole depths. This 
natural observation turns out to be correct and the skyrmions, for small values of the hole depth $\Gamma$ and distance of 
skyrmion separation $d$, do behave as a bound state. Tables \ref{d=4 hole table} and \ref{d=5.5 hole table} demonstrate the 
variation in the binding energy for the skyrmion configuration with the hole depth, for a fixed value of the width $b=2$. The data 
shows that as the hole depth gets larger, the binding energy increases. As we will discuss in the later sections, this method of analysis 
does not include the effect of the hole on the skyrmion during its motion. Moreover, the binding energy is more of a dynamic quantity. Skyrmions 
are extended objects and in an asymmetric system one of the skyrmions feels the effect of the hole more than the other. Although the energy of each 
skyrmion can vary, for small values of $\Gamma$ and $d$ the static binding energy provides a good explanation of the behaviour of the skyrmions.

\begin{table}

 \begin{center}
\begin{tabular}{|c|c|c|c|}

\hline
$|\Gamma|$ &  $E_{2}/8\pi$  & $E_{1}/8\pi$ & $E_{B}/8\pi $ \\
\hline \hline
0.1 & 1.9189 & 1.0662 & -0.2135 \\
0.2 & 1.9117 & 1.0630 & -0.2143 \\
0.3 & 1.9046 & 1.0598 & -0.2150 \\
0.4 & 1.8975 & 1.0566 & -0.2157 \\
0.5 & 1.8904 & 1.0534 & -0.2164 \\
\hline

\end{tabular}

\caption{\label{d=4 hole table} Table showing the variation in the total energy and binding energy for a $d=4$ 
two-skyrmion configuration interacting with a potential hole of width $b=2$ for various values of $\Gamma$.}
\end{center}

\end{table}

%
%
%
%
%
%

\begin{figure}[p]
\unitlength1cm \hfil

\begin{picture}(13,13)
 \epsfxsize=10cm \put(0,8){\epsffile{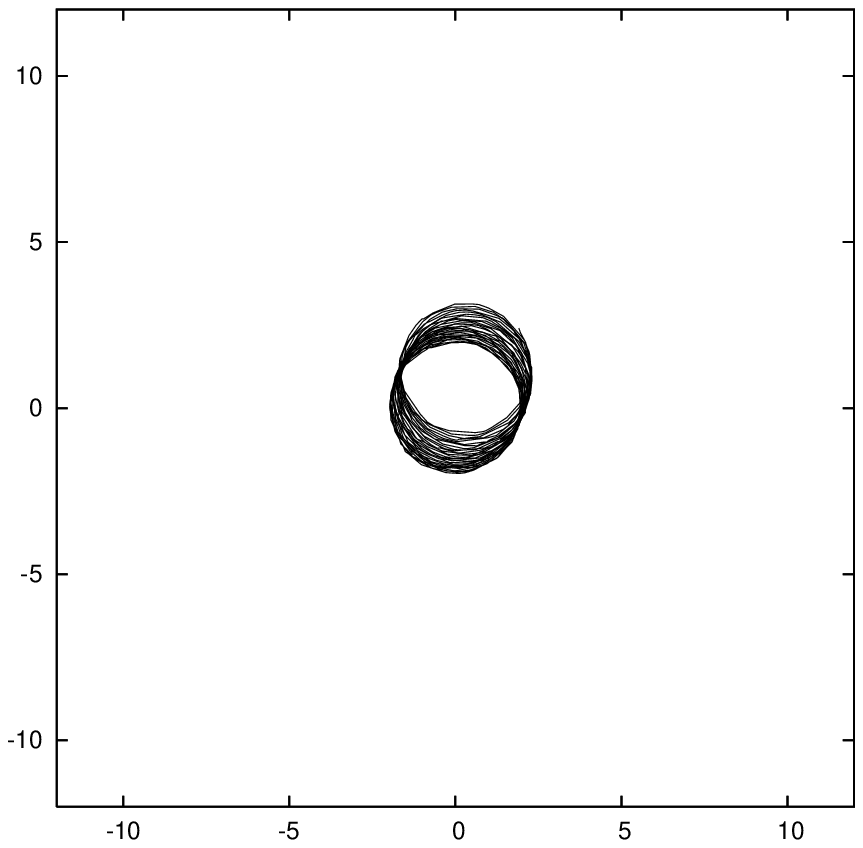}}
 \epsfxsize=10cm \put(7,8){\epsffile{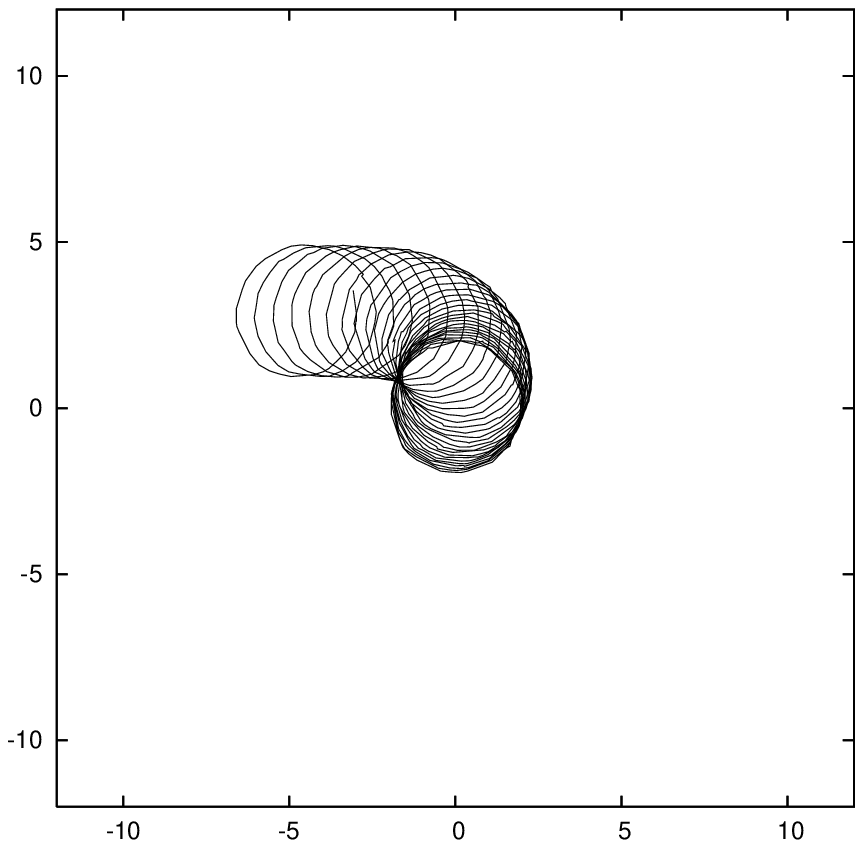}}
 \epsfxsize=10cm \put(0,0.25){\epsffile{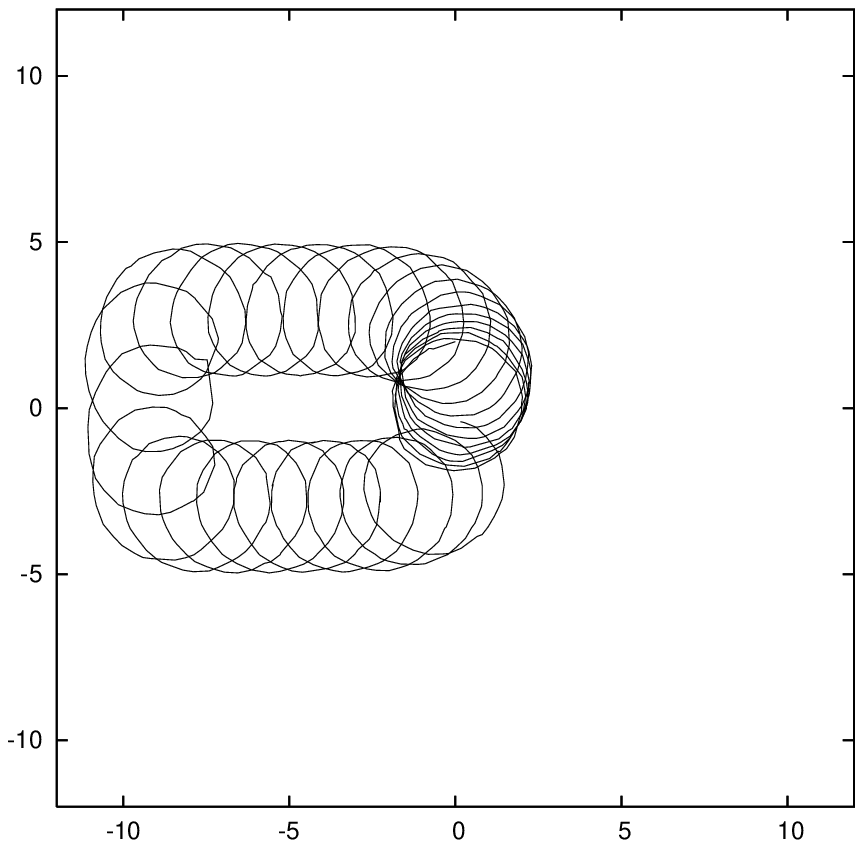}}
\epsfxsize=10cm \put(7,0.25){\epsffile{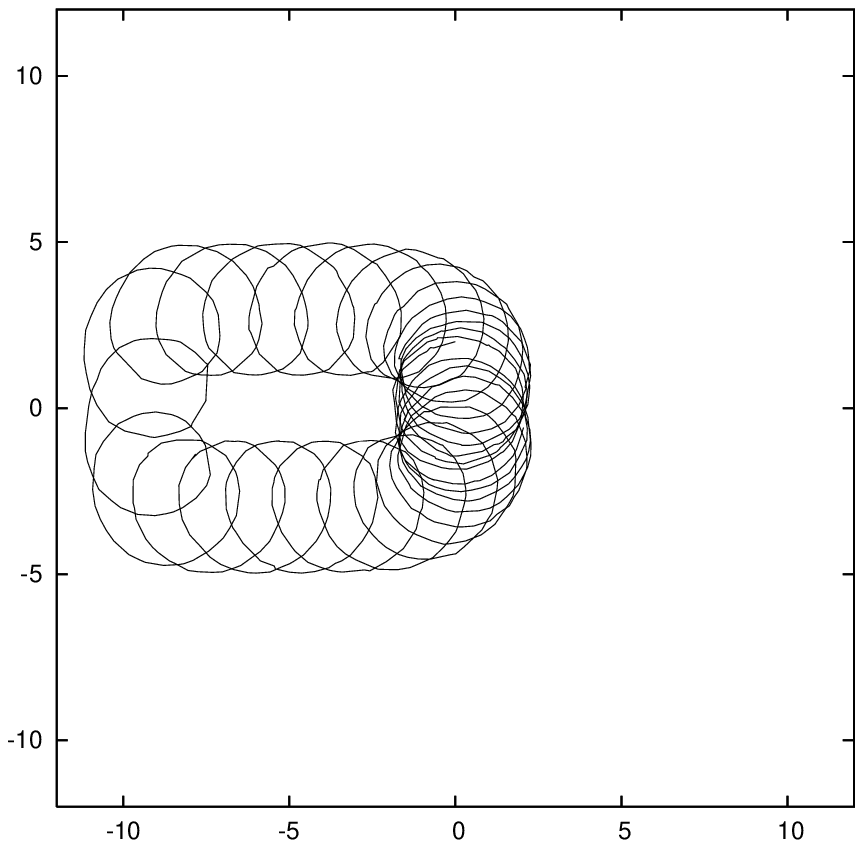}}

\put(3.75,0){c}
\put(10.75,0){d}
\put(3.75,7.5){a}
\put(10.75,7.5){b}

\end{picture}
\caption{\label{hole d=4 traj}Plots of the trajectories for the upper skyrmions in a system with an asymmetric potential 
hole of width $b=2$ with $d=4$ for various values of $\Gamma$: a) $\Gamma = -0.1$, b) $\Gamma = -0.2$, c) $\Gamma = -0.4$ and  d) $\Gamma = -0.5$, where the horizontal 
axis represents the $x$ coordinate and the vertical axis the $y$ coordinate. }

\end{figure}

%
%
%
%

Fig. \ref{hole d=4 traj} shows the trajectories of the upper skyrmion of a $d=4$ configuration interacting with a potential hole for various 
values of the hole depth $\Gamma$.  The trajectories show that the skyrmions are still bound, as was 
suggested by the binding energy analysis. Each plot was taken for a time length of $1000$ secs. The behaviour of the skyrmions in this system 
is reminiscent of a sling-shot type motion. Due to the small value of $d$, the skyrmions 
are tightly bound. When the skyrmions initially try to execute the normal circular motion, the upper skyrmion, $s_{1}$, is moving towards the 
hole and the lower, $s_{2}$, is moving into the space away from the hole. 
As $s_{1}$ approaches the hole its energy is reduced due to its tail interactions with the hole. To conserve the total energy of the system the 
skyrmions separate a little from each other. This separation increases as $s_{1}$ gets into the hole. In the meantime $s_{2}$ moves further away 
from the hole as $s_{1}$ approaches it. Classically we know that objects in a conserved system should speed up as they traverse a potential hole. 
Thus as $s_{1}$ enters the hole it quickly traverses it. The interaction between the two skyrmions during this period results in $s_{2}$ being 
sling shot around the hole towards the boundary at infinity. Once both skyrmions are out of the hole the normal motion can resume and the distance 
between the skyrmions corrects itself, resulting in $s_{2}$ being kept to the left of the hole, as we observe it. This motion continues as the 
skyrmions move along the edge of the hole. The plots corresponding to larger values of $|\Gamma|$ show that the skyrmions have travelled further 
than for the smaller 
values of $|\Gamma|$, over the same time length. This is due to the hole depth, where for a larger value of $|\Gamma|$ the skyrmions 
move faster across the hole, resulting in the increase in the distance travelled in the same amount of time. 

Although not shown here, the skyrmion configuration for $d=5$ produces similar plots to Fig. \ref{hole d=4 traj}, with a few differences.
For a $d=5$ system the skyrmions, although still executing circular 
motion about the configuration centre, follow a trajectory that is more distorted than the skyrmions of Fig. \ref{hole d=4 traj}. The 
skyrmions in this system are also able to penetrate the hole more than the skyrmions in the $d=4$ case. The ability to penetrate the 
hole is due to the skyrmions' binding energy. As before, let $s_{1}$ be the upper skyrmion and $s_{2}$ the lower. The skyrmions try to 
execute their normal circular motion but as $s_{1}$ approaches the hole its energy decreases. The initial binding 
energies for the $d=4$ and $d=5$ system are largely different due to the attractive channel in which they were created. As the energy of 
one skyrmion decreases the skyrmions separate from each other to conserve energy. Consequently, the binding energy or the interaction 
between the skyrmions, is reduced. In the system for $d=4$, the skyrmions do not separate as much from each other as in the $d=5$ case 
due to the tightness of the bound state. In the $d=5$ state the initial maximum 
separation between the skyrmions occurs  at the point when $s_{1}$ is in the hole. The weak interaction between the skyrmions slows down the distorted  rotation about the configuration centre. 
 In comparison with the trajectories of Fig. \ref{hole d=4 traj}, the skyrmions' rate of rotation is not as reduced 
by the hole as in the $d=5$ system. Since the skyrmions are much more tightly bound at the start, the effect of the hole upon the 
interaction between the skyrmions is much less pronounced. Therefore the skyrmion trajectory is less distorted and the skyrmions move faster while traversing the hole.

\begin{table}

 \begin{center}
\begin{tabular}{|c|c|c|c|}

\hline
$|\Gamma|$ &  $E_{2}/8\pi$  & $E_{1}/8\pi$ & $E_{B}/8\pi $ \\
\hline \hline
0.1 & 2.1099 & 1.0683 & -0.0267 \\
0.2 & 2.1058 & 1.0673 & -0.0288 \\
0.3 & 2.1017 & 1.0663 & -0.0309 \\
0.4 & 2.0977 & 1.0652 & -0.0327\\
0.5 & 2.0936 & 1.0642 & -0.0348\\
\hline

\end{tabular}

\caption{\label{d=5.5 hole table} Table showing the variation in the total energy and binding energy for a $d=5.5$ 
two-skyrmion configuration interacting with a potential hole of width $b=2$ for various values of $\Gamma$.}
\end{center}

\end{table}








\subsection{Transition dynamics}

\begin{figure}[p]
\unitlength1cm \hfil

\begin{picture}(13,13)
 \epsfxsize=10cm \put(0,8){\epsffile{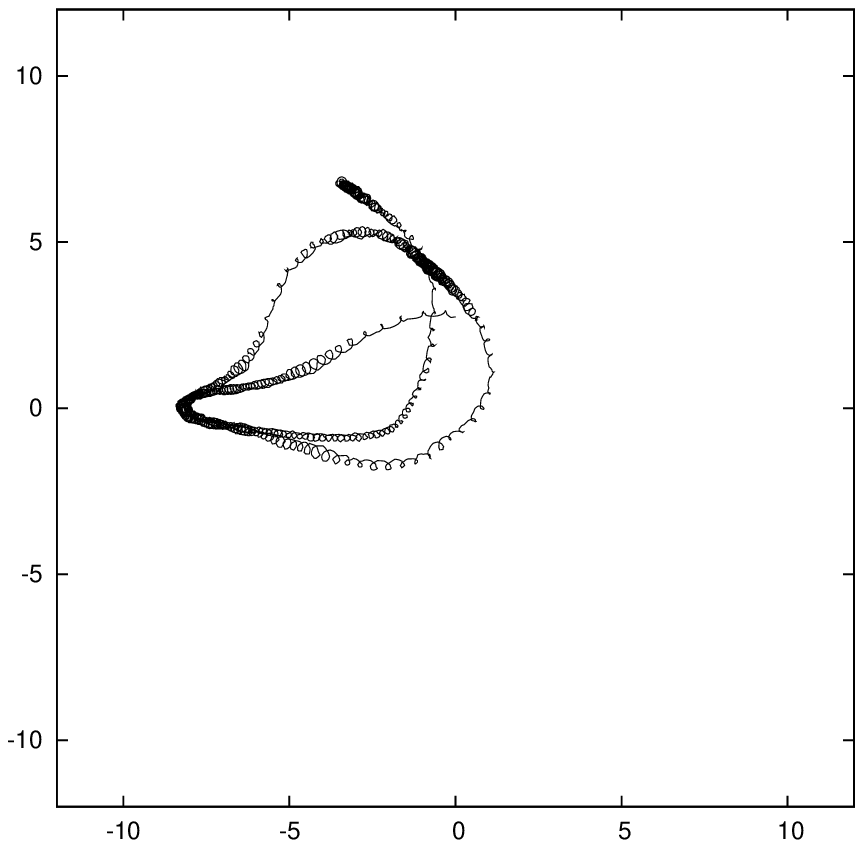}}
 \epsfxsize=10cm \put(7,8){\epsffile{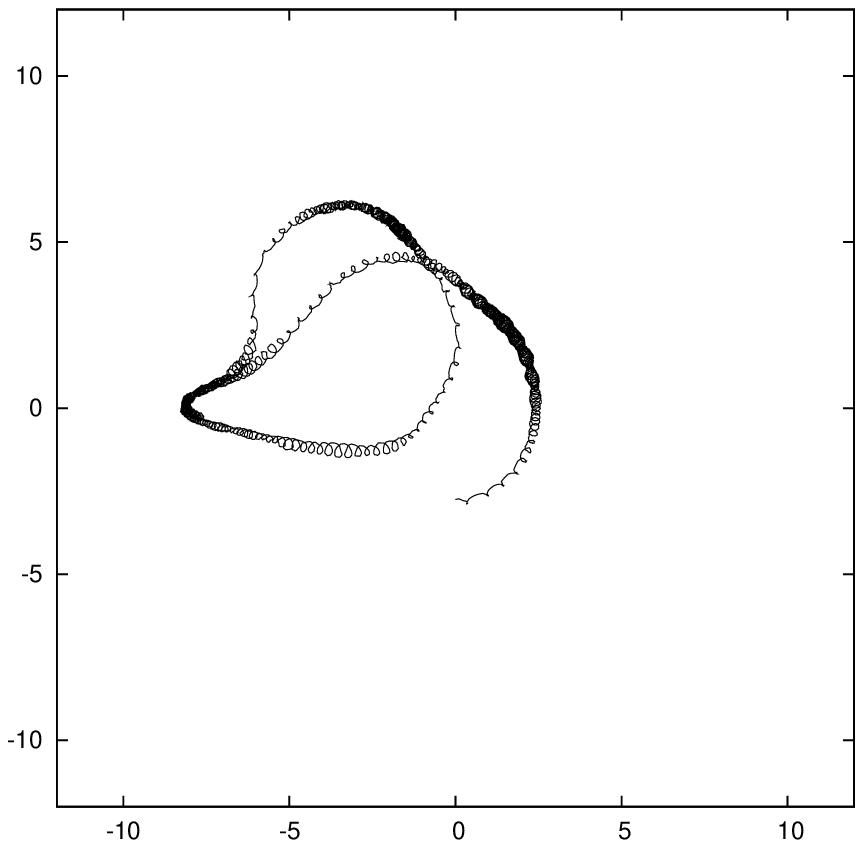}}
 \epsfxsize=10cm \put(0,0.25){\epsffile{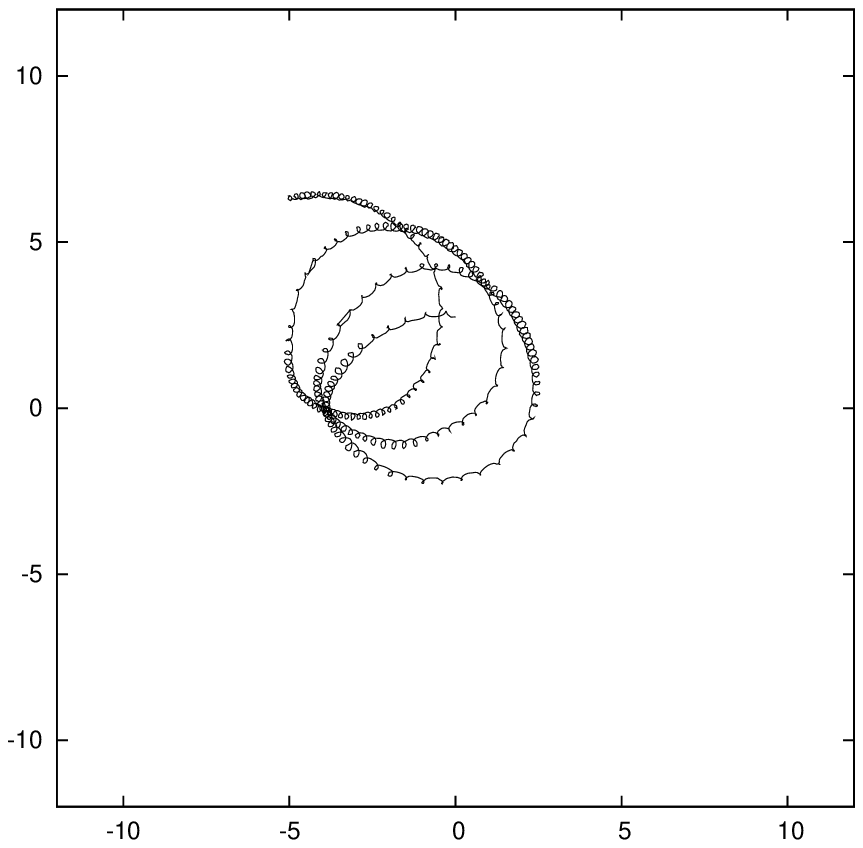}}
\epsfxsize=10cm \put(7,0.25){\epsffile{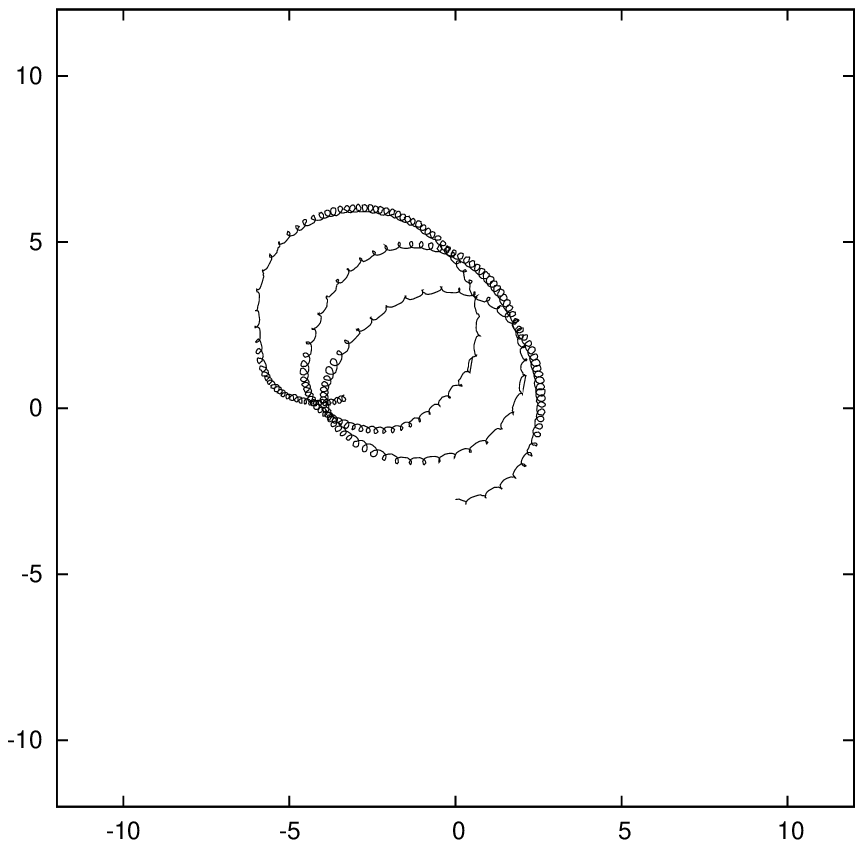}}
\put(5,0){c}
\put(12,0){d}
\put(5,7.5){a}
\put(12,7.5){b}

\end{picture}
\caption{\label{hole trans 0.2,0.3} Plots of the trajectories for the upper and lower skyrmions of a system with 
an asymmetric potential hole of width $b=2$ with $d=5.5$ for various values of $\Gamma$: a) $\Gamma = -0.3$, upper  skyrmion , b) $\Gamma = -0.3$, lower skyrmion, 
c) $\Gamma = -0.2$, upper skyrmion and  d) $\Gamma = -0.2$, lower skyrmion, where the horizontal 
axis represents the $x$ coordinate and the vertical axis the $y$ coordinate.}

\end{figure}

\begin{figure}[p]
\unitlength1cm \hfil

\begin{picture}(13,13)
 \epsfxsize=10cm \put(0,8){\epsffile{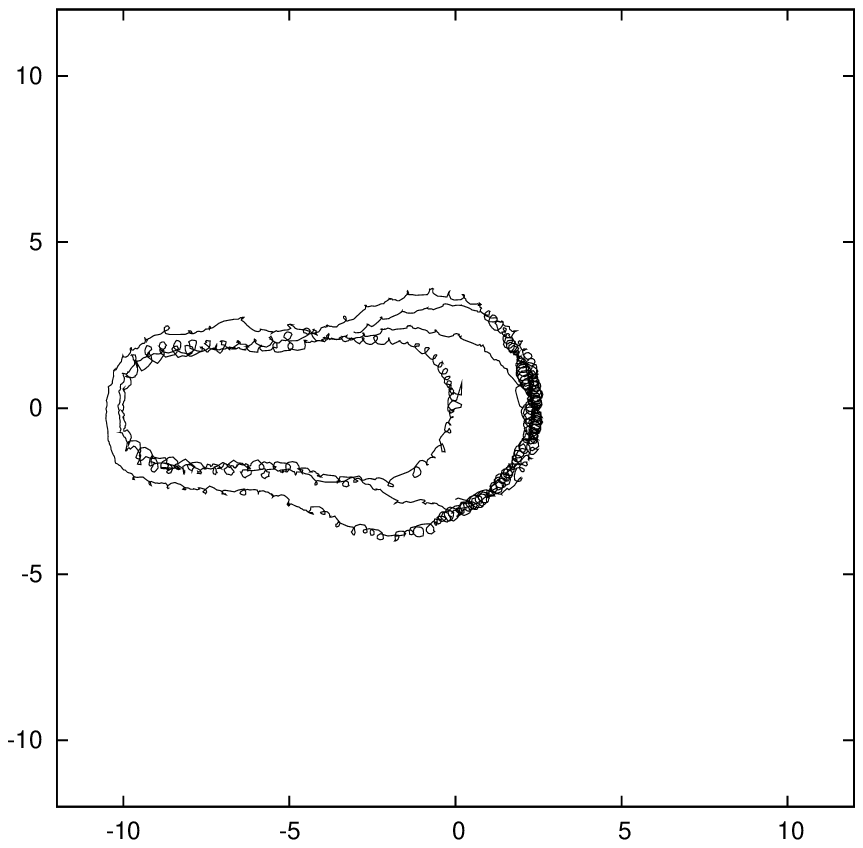}}
 \epsfxsize=10cm \put(7,8){\epsffile{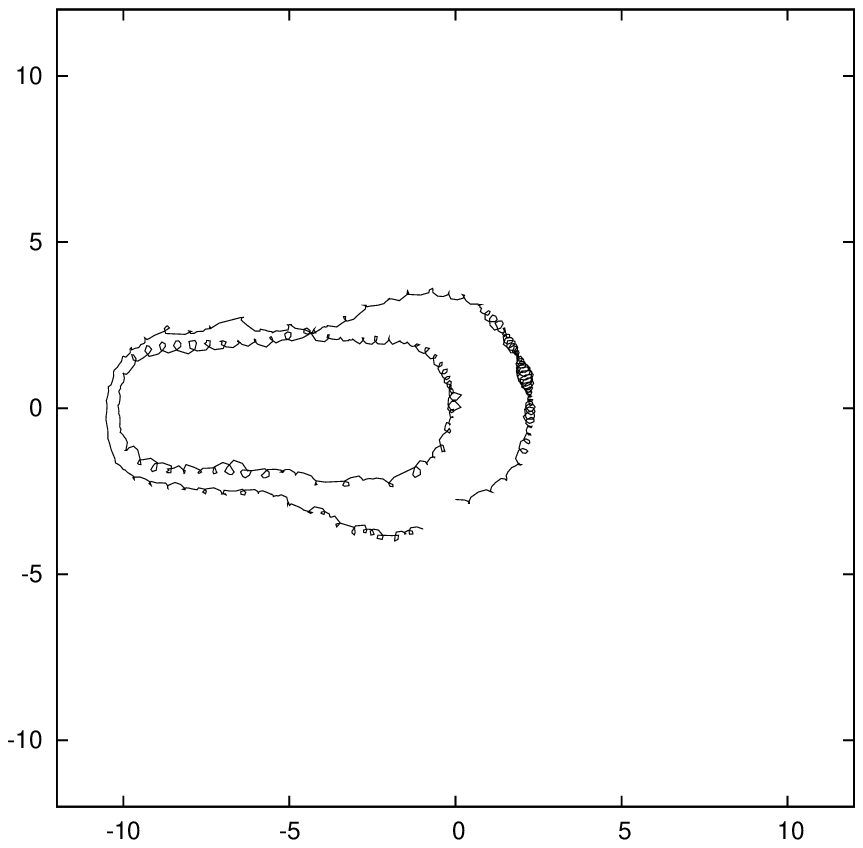}}
 \epsfxsize=10cm \put(0,0.25){\epsffile{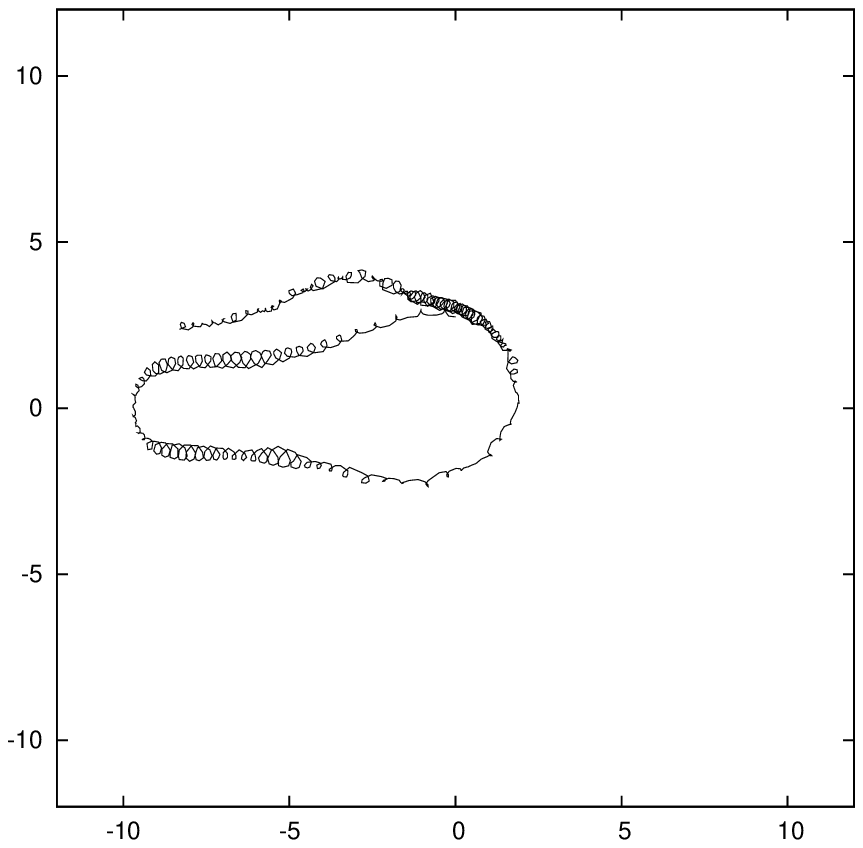}}
\epsfxsize=10cm \put(7,0.25){\epsffile{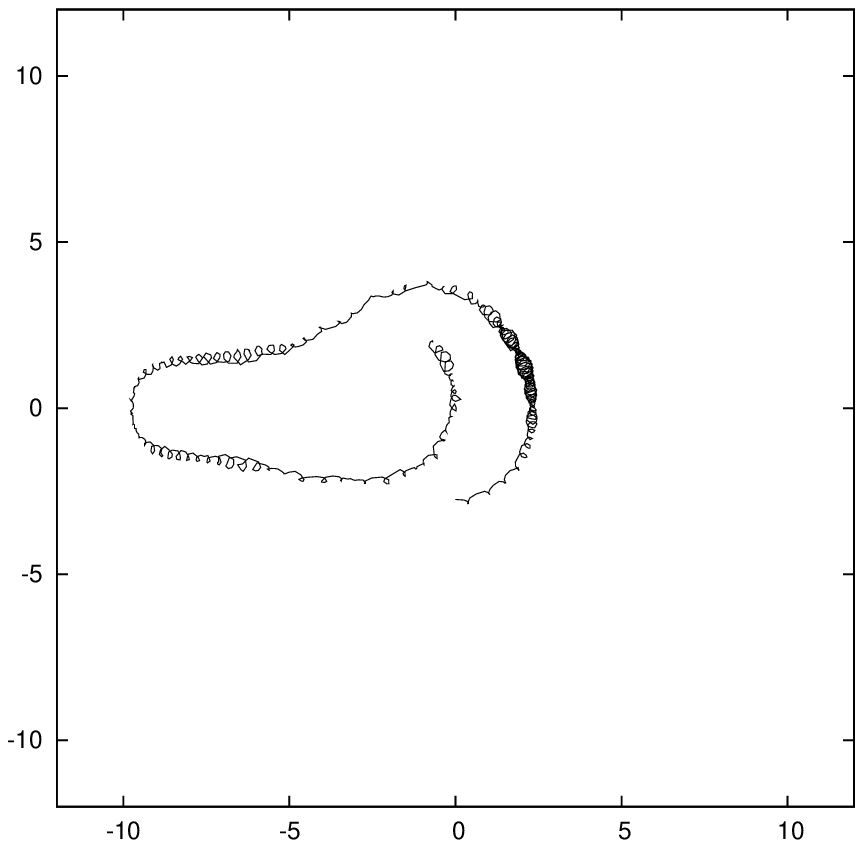}}
\put(5,0){c}
\put(12,0){d}
\put(5,7.5){a}
\put(12,7.5){b}

\end{picture}
\caption{\label{hole trans 0.4,0.5} Plots of the trajectories for the upper and lower skyrmions of a system with 
an asymmetric potential hole of width $b=2$ with $d=5.5$ for various values of $\Gamma$: a) $\Gamma = -0.5$, upper skyrmion, b) $\Gamma = -0.5$, lower skyrmion, c) $\Gamma = -0.4$, upper skyrmion and  d) $\Gamma = -0.4$, lower skyrmion, where the horizontal 
axis represents the $x$ coordinate and the vertical axis the $y$ coordinate.}

\end{figure}
In this section we describe the transition from the bound systems described in the previous section to one in which the skyrmions behave 
as if they were unbound, similar to the dynamics seen in the symmetric system of \cite{CZ09}. The transition state is best observed through 
the variation in $\Gamma$ for the $ d=5.5$ system. It can be seen from the trajectories in Fig. \ref{hole trans 0.2,0.3} that the skyrmions 
for $\Gamma=-0.2$ still form a bound state. However, as we increase the hole depth from $\Gamma=-0.2$ to $\Gamma=-0.3$ and further to 
$\Gamma=-0.5$, the dynamics is no longer bound and the skyrmion trajectories are completely uncorrelated, as seen in Fig. \ref{hole trans 0.4,0.5}.
 In Fig. \ref{hole trans 0.2,0.3}a and b the 
skyrmions are able to penetrate the hole more than in other previous systems. As one skyrmion penetrates the 
hole the other remains stationary outside. The skyrmions in this state are just on the point of being broken up by the hole. The skyrmions 
still remain bound after the upper skyrmion comes out of the hole and this is evident in the trajectory. In the case of $\Gamma=-0.4,-0.5$, 
the skyrmions behave differently. Here one of the skyrmions moves along the axis of the hole after a small initial circular motion. The other comes to rest after 
this initial movement. The other skyrmion continues along its path until it reaches the boundary, where it gets reflected and moves back 
towards the stationary skyrmion. When they get close to each other, the skyrmions violently oscillate due to their mutual interaction and these oscillations are discussed
in a later section.

%
%
%
%
%

Let us consider the binding energy, $E_{B}$, defined in (\ref{EB}) to be a function of $D$, the distance a single skyrmion
is from the asymmetric hole in (\ref{EB}). By examining the binding energy for the skyrmions in these systems as a function of $D$, it can be seen how the skyrmions 
become unbound as they get close to the hole.  By examining tables \ref{d=4 hole table} and  \ref{d=5.5 hole table} and utilising other information, 
we can examine how the single skyrmion energy varies with $D$. 
The energies of the two skyrmion configuration in the presence of a hole of depth $\Gamma= -0.4$ is $E_{2}=2.0977/ 8\pi$ and for $\Gamma= -0.5$ 
is $E_{2}=2.0936/ 8\pi$. 
Tables \ref{gam_04 hole trans} and \ref{gam_05 hole trans} show how the binding energy of the skyrmion configuration varies when one of the 
skyrmions approaches the hole. It is clear that when the skyrmions, in the case of $\Gamma=-0.4$, get near $D=1$ the binding energy of their 
configuration becomes positive and the skyrmions are no longer bound. Thus as the skyrmion approaches the hole, after the initial circular 
motion, its starts to feel the effect of the hole. To conserve energy the skyrmions must increase their distance of separation. In order to 
do this the upper skyrmion moves along the edge of the hole. The more the skyrmion feels the hole the more the configuration separates.  
As the binding energy increases the skyrmions interact less and the upper skyrmion is free to move along the edge of the hole towards the
boundary. After the reflection from the boundary the skyrmion speeds up traversing the obstruction, moving towards the opposite edge of the
 hole. This process then repeats itself; when the skyrmion reaches the stationary one it interacts with it causing the the stationary skyrmion 
to begin moving. The previously moving skyrmion now comes to rest. It was evident in the previous section that as $\Gamma$ increases, for a given value of $d$, 
the skyrmions' trajectories change. The skyrmions cease to behave as a  bound state and the behaviour of the two skyrmions becomes uncorrelated.

\begin{table}

 \begin{center}
\begin{tabular}{|c|c|c|c|}

\hline
$D$ &  $E_{1}/8\pi$ & $E_{B}/8\pi $ \\
\hline \hline
2.5 & 1.0623 & -0.0269 \\
2.0 & 1.0566 & -0.0155 \\
1.5 & 1.0503 & -0.0029 \\
1.0 & 1.0458 & +0.0061 \\
0.5 & 1.0432 & +0.0072\\
\hline

\end{tabular}

\caption{\label{gam_04 hole trans} Table showing the variation in the binding energy for a $\Gamma=-0.4$ 
hole of width $b=2$ for various values of the single skyrmion distance from the hole $D$.}
\end{center}

\end{table}

\begin{table}

 \begin{center}
\begin{tabular}{|c|c|c|c|}

\hline
$D$ &  $E_{1}/8\pi$ & $E_{B}/8\pi $ \\
\hline \hline
2.5 & 1.0613 & -0.029 \\
2.0 & 1.0534 & -0.0132 \\
1.5 & 1.0455 & +0.0026 \\

\hline

\end{tabular}

\caption{\label{gam_05 hole trans} Table showing the variation in the binding energy for a $\Gamma=-0.5$ hole of 
width $b=2$ for various values of the single skyrmion distance from the hole $D$.}
\end{center}

\end{table}

The initial distance of closest approach of the upper skyrmion can be estimated from our dynamic binding energy arguments. In the $\Gamma=-0.4$ 
system, the closest distance the skyrmion reaches, based upon the trajectory, is approximately $D=1.3$. Comparing this with table \ref{gam_04 hole trans}, 
we can see that the transition point occurs between $D=1.5$ and $D=1$. Based upon the numerical values of each, the transition point is closer to $1.5$ 
than $1$.  In the $\Gamma=-0.5$ case, the closest distance is about $D=1.75$, based upon the trajectory. Estimating this distance from the energetics, 
we see that the transition point occurs between $D=2$ and $D=1.5$, with it being 
nearer to $1.5$. Thus from just energetic arguments we have a reasonable estimate of the closest distance the skyrmion can initially get to the hole.

In the section on the potential barrier system we examined the effect of increasing the bounding box of the simulation. It was found that the skyrmions trajectory away 
from the obstruction was uniquely determined by the boundary. In the potential hole system the boundary once again determines the dynamics of the skyrmions. The skyrmions in such a system 
often reach the boundary where they get reflected. If identical skyrmions are put into a system with a larger grid, the skyrmions undergo a reflection from the boundary 
at a later time than for a system of a smaller grid. 





\section{Angular Momentum } \label{ang mom and oscillatons}

In this current study we have considered the scattering of a skyrmion configuration off an asymmetric obstruction, similar to the symmetric obstructions of \cite{CZ09},
except now the obstruction no 
longer exists for $x >0$. In this section we shall derive the contributions the potential obstructions make to the derivative of the orbital 
angular momentum $\dot{l}$. We shall also discuss the trends seen in the data obtained for the orbital angular momentum, $l$. It will be seen 
that the time variation in $l$ can be decomposed into the behaviour of the average size of the skyrmions, $r$, and the modulus of the skyrmion 
guiding centre, $\underline{R}$, both defined previously. It was noted in the potential hole section that the simulations of the skyrmions in the transition dynamics exhibit
oscillations in the energy density. Here we shall show that these oscillations play a significant role in the conservation of the orbital angular momentum.

\subsection{Asymmetric potential obstructions contribution to $\dot{l}$}\label{asym int}

 The potential obstructions are introduced as an inhomogeneity in the potential term's, $V(\underline{\phi})$, 
coefficient $\gamma_{3}$. The obstructions can be defined more concisely if they are written  using Heaviside functions.
In this formalism we can rewrite the approximation of the obstruction as:
\begin{equation}
V(\underline{\phi})=\frac{1}{2} ( 1-\phi_{3}^{2} )  + \frac{1}{2} \Gamma  ( 1-\phi_{3}^{2} )  \Big\{  \Theta(-x + x_{0})  \big[ \Theta(y+y_{0})-\Theta(y-y_{0}) \big] \Big\} \nonumber .
\end{equation}
We now proceed, as in \cite{CZ09}, to calculate the contribution the obstruction makes to $\dot{q}$: 
\begin{eqnarray}
{\dot {q}_{obs}}&=&-\epsilon_{ij}\partial_{i}\left ( \frac{\delta V(\underline{\phi})}{\delta \underline{\phi}} \cdot{\partial_{j} \underline \phi} \right )  \nonumber \\
&=& -\epsilon_{\mu\nu}\partial_{\mu}\Big\{ -\phi_{3}\partial_{\nu}\phi_{3}   \big( 1 + \Gamma  \Theta(-x + x_{0}) \Big[   
\Theta(y+y_{0})-\Theta(y-y_{0})  \Big]  \big)  \Big\}. \nonumber
\end{eqnarray}
The time derivative of the orbital angular momentum was defined previously by (\ref{ldot}). If the above is inserted into (\ref{ldot}), then an expression 
for the contribution that the asymmetric potential obstruction makes to $\dot{l}$ can be found:
\begin{eqnarray}
\dot {l}_{obs} &=&\frac{1}{2} \int_{\mathbb{R}^{2}}d^{2}x \, \underline{x}^{2}\dot {q}_{obs}\nonumber \\
&=& - \frac{\Gamma}{2}\int_{\mathbb{R}^{2}}d^{2}x \, \underline{x}^{2}\partial_{y}(\frac{1}{2}\phi_{3}^{2}) \bigg (  \delta(-x+x_{0})  
\Big[ \Theta(y+y_{0})-\Theta(y-y_{0})\Big]  \bigg ) \nonumber \\
&  +&\frac{\Gamma}{2}\int_{\mathbb{R}^{2}}d^{2}x \, \underline{x}^{2}\partial_{x}(\frac{1}{2}\phi_{3}^{2}) \Theta(-x+x_{0}) \delta(y+y_{0})  \nonumber \\  
&  -&\frac{\Gamma}{2} \int_{\mathbb{R}^{2}}d^{2}x \,\underline{x}^{2} \partial_{x}(\frac{1}{2}\phi_{3}^{2}) \Theta(-x+x_{0}) \delta(y-y_{0}) . \nonumber 
\end{eqnarray}

These integrals can be easily computed to give the overall contribution of the asymmetric obstruction to $\dot{l}$, given by:
\begin{eqnarray}
\dot {l}_{obs} &=&\frac{ \Gamma}{2} \int_{-\infty}^{x_{0}} dx \, \underline{x}^{2}\partial_{x}(\frac{1}{2}\phi_{3}^{2}) \Bigg |_{y=-y_{0}} 
- \frac{ \Gamma}{2} \int_{-\infty}^{x_{0}} dx \, \underline{x}^{2}\partial_{x}(\frac{1}{2}\phi_{3}^{2}) \Bigg |_{y=y_{0}}  \nonumber  \\
&+&\frac{ \Gamma}{2}\int_{y_{0}}^{\infty} dy \, \underline{x}^{2} \partial_{y}(\frac{1}{2}\phi_{3}^{2})  \Bigg |_{x=x_{0}}   
- \frac{ \Gamma}{2}\int_{-y_{0}}^{\infty} dy \,  \underline{x}^{2} \partial_{y}(\frac{1}{2}\phi_{3}^{2}) \Bigg |_{x=x_{0}}  \\
&=&\frac{1}{2}\int_{\mathbb{R}^{2}}d^{2}x \,\underline{x}^{2}\dot {q}_{obs} . \nonumber
\end{eqnarray}

\subsection{Examination of $J$ and $\dot{J}$ }

\begin{figure}[p]

\unitlength1cm \hfil
\begin{picture}(14,14)

\epsfxsize=10cm \put(0,0.25){\epsffile{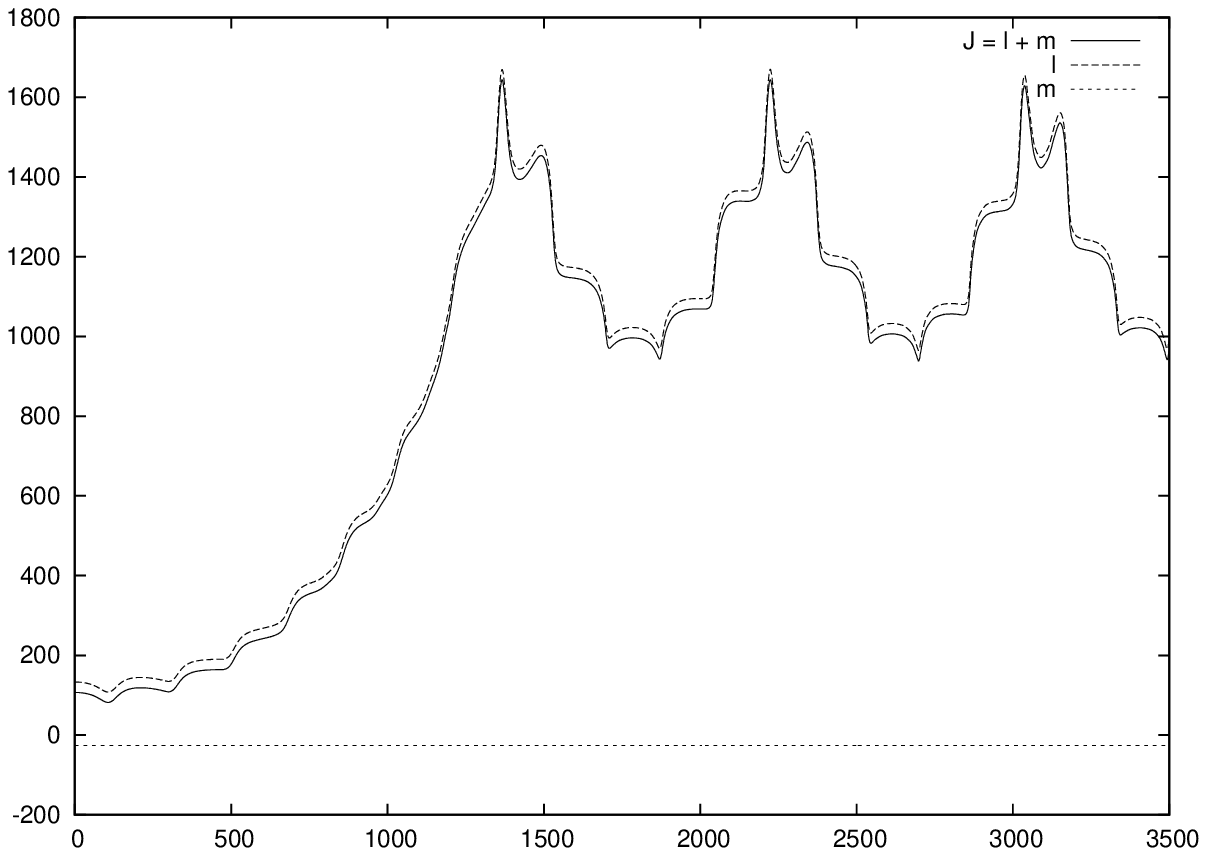}}
\epsfxsize=10cm \put(0,8.25){\epsffile{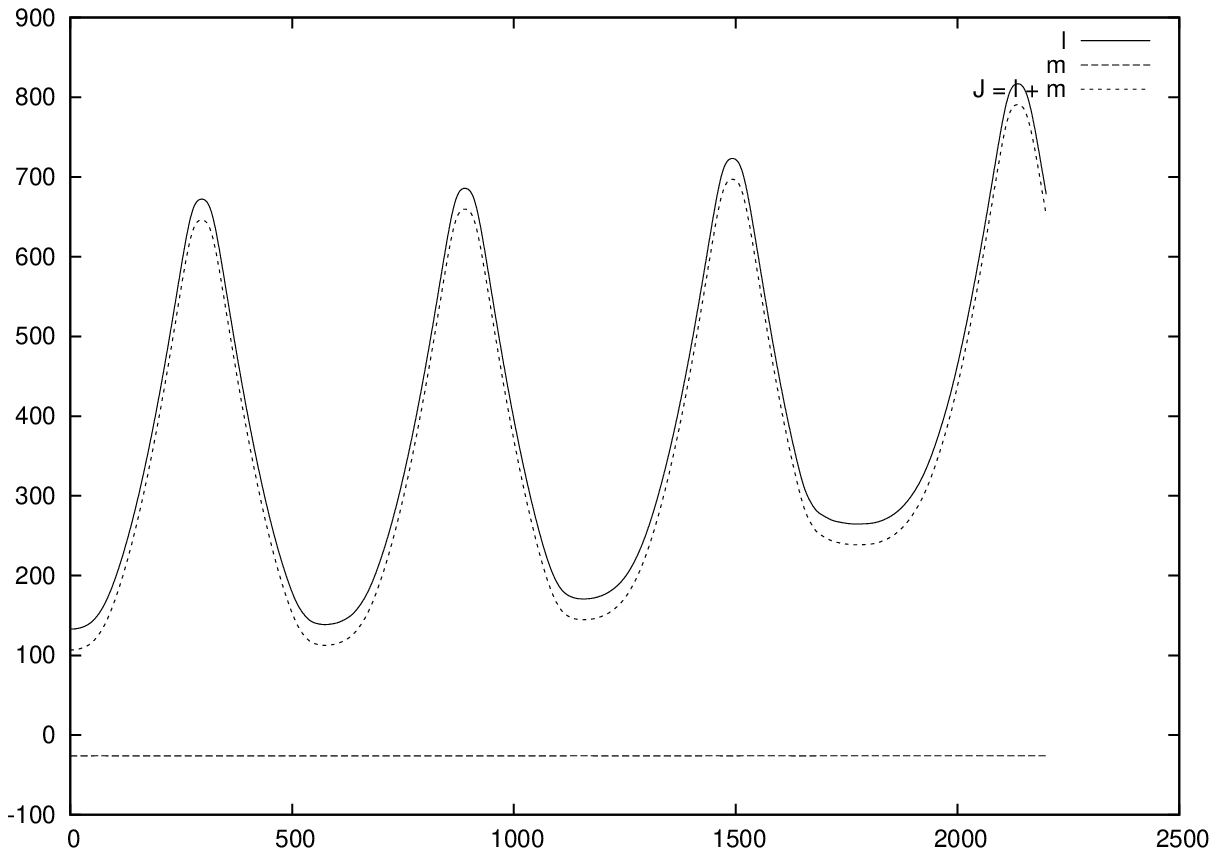}}

\put(5.25,7.75){a}
\put(5.25,0){b}

\end{picture}
\caption{\label{l barrier and hole} Plots of the total angular momentum, $J=l+m$, and its components as a function of time for a $d=6$ skyrmion configuration 
interacting with a potential hole or barrier for $\Gamma= \pm 0.25$: a) hole and  b) barrier.}
\end{figure}

In any system the total angular momentum, $J=l+m$,  should be conserved. However, it was discovered in \cite{CZ09} that for scattering processes 
in Landau-Lifshitz models the total angular momentum, as defined by (\ref{l}), is not conserved. On further examination it was found that throughout
 all the simulations, the total magnetization in the third direction, $m$, is well conserved and it is the orbital angular momentum, $l$, which is not 
conserved with time. In the asymmetric system such a situation presents itself again, where the total angular momentum 
is not conserved due to the non-conservation of the orbital angular momentum (\ref{l}), denoted $\dot{l}_{fields}$.
The non-conservation of $J$ is apparent in the plots of Fig. \ref{l barrier and hole}. Both of these plots show
the total angular momentum, $J=l+m$ and its components, for a $d=6$ skyrmion configuration, interacting with a potential hole or barrier with $\Gamma=\pm0.25$. 
Here it is evident that for either a hole or a barrier the total
angular momentum is not conserved due to the non-conservation of $l$, since $m$ is well conserved throughout. Before analysing this 
apparent non-conservation
of $J$ we shall firstly discuss the general behaviour of $l$ and hence $J$, during the simulations.

The time evolution of $l$, as formulated by Papanicolaou and Tomaras \cite{PT91}, 
provides us with an insight into the behaviour of the skyrmions during the simulations. It was already seen that  
the average skyrmion size, $r$ and the guiding centre coordinate, $\underline{R}$, can be related to the orbital angular momentum, $l$, by :
\begin{equation}
r^{2}= \frac{l}{2 \pi Q}- \underline{R}^{2}, \nonumber
\end{equation}
or 
\begin{equation}
l=2\pi Q \{ r^{2}+ \underline{R}^{2}\}. \nonumber
\end{equation}

 The guiding centre coordinate, $\underline{R}$, 
for  a single skyrmion system provides the location
of the skyrmion centre. In the case of a two-skyrmion configuration, $\underline{R}$ provides the position of 
the centre of the configuration. The initial placement of the skyrmions
results in the centre of the configuration nearly positioned at (0,0). In the case of a hole or barrier, the asymmetry of the system results in the position of
the configuration centre becoming a function of time. How it varies during the simulation depends on the system. 
In the case of a potential barrier the 
skyrmions collectively move about the system.  Therefore as the skyrmions begin to move, $\underline{R}$ deviates from $(0,0)$ 
and $|\underline{R}|$ steadily increases. $|\underline{R}|$ continues 
to increase until the skyrmions reach the boundary of the system. As the skyrmions orbit the boundary, $|\underline{R}|$ slowly increases and decreases as the skyrmions move 
towards and away from the boundary. If we consider the behaviour 
of $l$ as corresponding to the dual behaviour of the quantities $r$ and $\underline{R}$, then as $|\underline{R}|$ steadily increases, so does $l$. 
This overall trend in $|\underline{R}|$ is evident 
in Fig. \ref{l barrier and hole}a  which clearly shows $l$ steadily increasing with time. 
In the case of a potential
hole we can use similar analysis. Fig. \ref{l barrier and hole}b shows a plot of the orbital angular momentum for potential hole with $\Gamma=0.25$.
 Here the increase in $l$, is much larger than in
the case of the barrier
system. In this particular system, the configuration centre moves much more than the barrier system and is periodic. This is evident if we recall the dynamics of 
the skyrmions in this system. Here one of the skyrmions moves along the edge of the hole while the other remains stationary.
Thus the centre of the configuration moves away from $(0,0)$ as the skyrmion approaches the boundary. After the skyrmion is reflected, $\underline{R}$ 
begins to approach $(0,0)$ once again. However, due to the dynamics $\underline{R}$ does not reach $(0,0)$ again. This periodic increase and decrease in 
$|\underline{R}|$ matches the behaviour of $l$ for the potential hole.

 The behaviour of the
orbital angular momentum, with $\underline{R}=0$, can be understood exactly in terms of the behaviour of $r$. The skyrmions tail is 
governed by value of the potential coefficient, 
$\gamma_{3}$ and the regions where this is modified means the skyrmion tails become modified. As the skyrmion moves into a region 
for which $\gamma_{3} \rightarrow 1 + \Gamma$, 
its tail contracts or expands. This continued shrinking and 
expansion, as the skyrmion moves on and off the potential obstructions, results in a time oscillation of $r^{2}$. In Fig. \ref{l barrier and hole}a 
the behaviour of $r$ is evident
in the time dependence of $l$. The periodic small oscillations occurring in $l$, correspond to the time variation of $r$. It was seen in the symmetric barrier system
that as the skyrmion configuration is `on' the obstruction, the skyrmions reduce their distance of separation. In the asymmetric barrier system the small oscillations in $l$
 are caused by this behaviour. The interplay between the time oscillation 
in $r^{2}$ and the gradual increase in $\underline{R}^{2}$  sufficiently accounts for the observed behaviour of the angular momentum.

\begin{figure}[htbp]

\unitlength1cm \hfil
\begin{picture}(14,14)

\epsfxsize=10cm \put(0,0.25){\epsffile{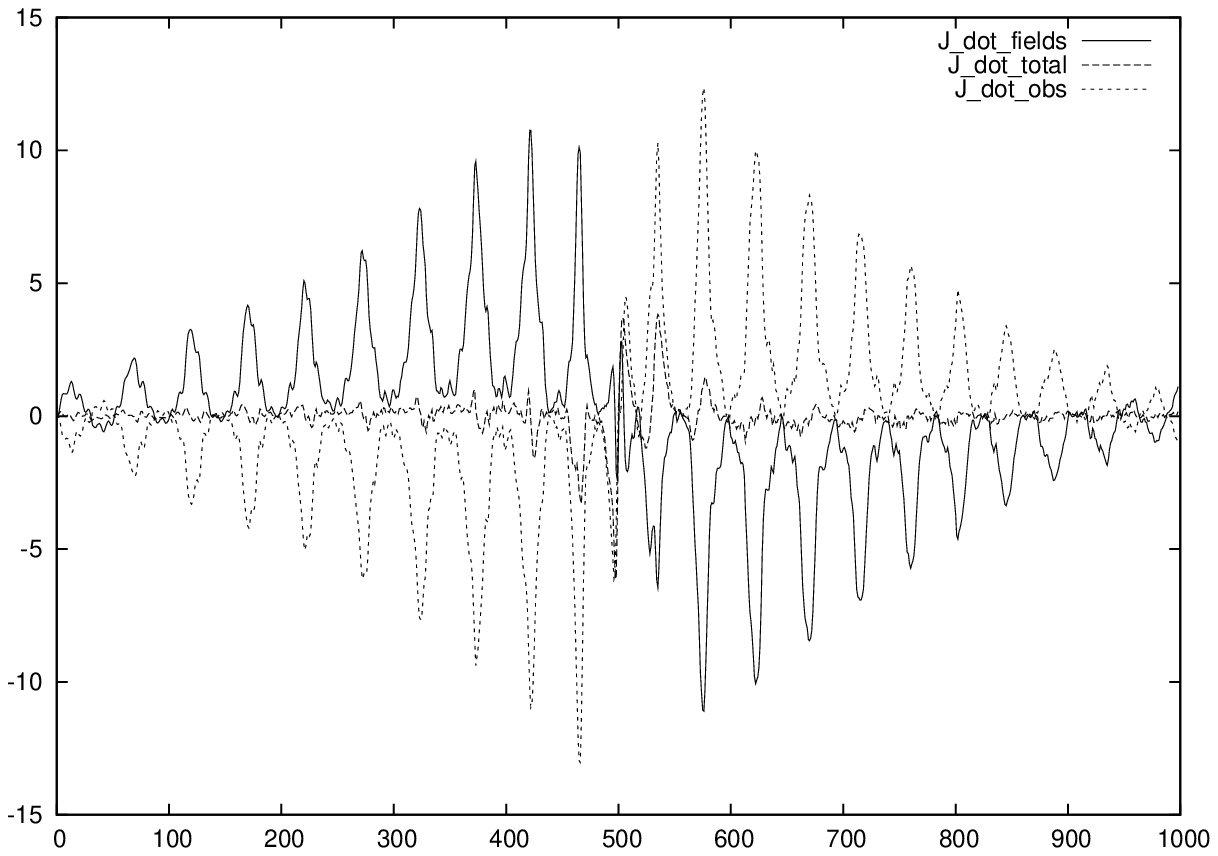}}
\epsfxsize=10cm \put(0,8.25){\epsffile{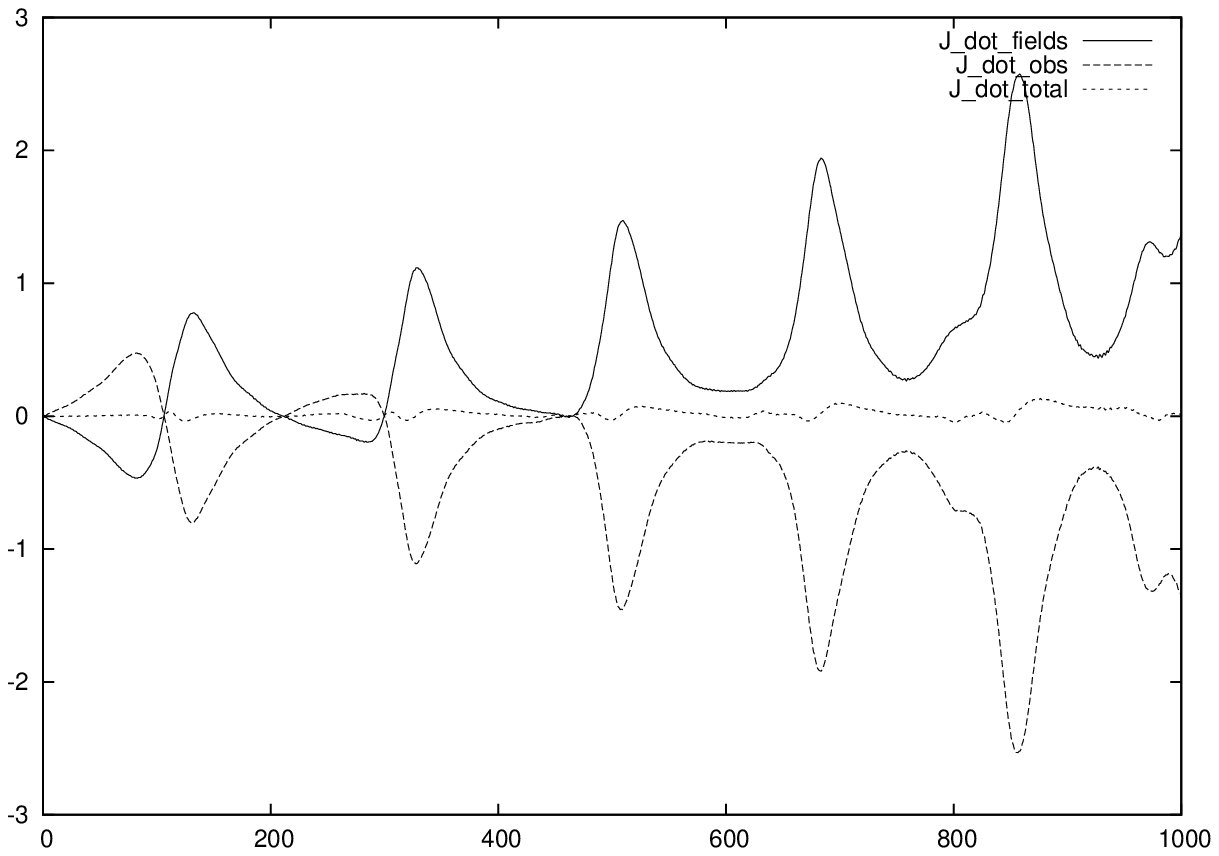}}

\put(5.25,7.75){a}
\put(5.25,0){b}

 \end{picture}
\caption{\label{l_dot compare h and b}Plots of the time derivative of the total angular momentum, $\dot{J}_{tot}$ and its components $\dot{l}_{fields}$ and $\dot{l}_{obs}$ as a function 
of time for a two different systems: a) $d=6$ skyrmion configuration with $b=2$ for $\Gamma= 0.25$ and b) $d=5$ skyrmion configuration with $b=2$ for $\Gamma= -0.5$}

\end{figure}

It has already been noted that the total angular momentum, $J$, is not conserved. It was discovered in our previous work that 
by examining how the potential obstructions were constructed, an analytical expression for the 
contribution the potential obstructions make to the time derivative of the orbital angular momentum, $\dot{l}$, was found. These expressions were in the form of two integrals 
which were calculated during the simulations. When the contribution of the potential obstructions was included in the expression for $\dot{J}$, the 
conservation of total angular momentum was restored. These same integrals, for the asymmetric 
obstruction, were computed in section \ref{asym int}
 and were calculated throughout the simulations. The contribution the obstruction makes to the derivative of the total 
angular momentum, $\dot{J}_{tot}$, is denoted $\dot{l}_{obs}$. The contribution to $\dot{J}_{tot}$ from the fields, given by (\ref{l}), is denoted $\dot{l}_{fields}$. 
Therefore the overall conservation of the total angular momentum can be written:
\begin{equation}
\dot{J}_{tot}= \dot{l}_{obs}+\dot{l}_{fields}, \nonumber
\end{equation} 
where there is no contribution due to the total magnetization in the third direction, $m$, since throughout $\dot{m}=0$.
Fig. \ref{l_dot compare h and b} shows the plots of the total angular momentum, $\dot{J}_{tot}= \dot{l}_{obs}+\dot{l}_{fields}$ for two different systems.
In Fig. \ref{l_dot compare h and b}a, $\dot{l}_{obs}$, $\dot{l}_{fields}$ and $\dot{J}_{tot}$ are plotted for a $d=6$ skyrmion configuration
interacting with a potential barrier of width $b=2$ and height $\Gamma=0.25$. It can seen from the plot that the overall behaviour of
 $\dot{l}_{fields}$ and $\dot{l}_{obs}$
match, resulting in $\dot{J}_{tot}$ being well conserved in time. It can be seen in the latter stages of the plot that this
conservation is slightly destroyed. The skyrmions at this point start to appreciably interact with the boundary of the system. It is due to
these boundary effects that $\dot{J}_{tot}\ne 0$ at these points. Fig. \ref{l_dot compare h and b}b shows the corresponding plot 
for a $d=5$ skyrmion configuration interacting with potential hole of depth $\Gamma=-0.5$. It is evident here that the behaviour of $\dot{l}_{fields}$ and $\dot{l}_{obs}$
is again nearly identical, resulting in the overall conservation of $J_{tot}$. We note again that there exist regions in Fig. \ref{l_dot compare h and b}b 
where $J_{tot}$ is not conserved. In this particular system these regions correspond to the points when one of the skyrmions
is reflected from the boundary and constitute another boundary effect.

\subsection{Skyrmion oscillations}

It was seen in the previous section that the total angular momentum, ${J}_{tot}$, for the asymmetric system is well conserved, just as it was 
for the symmetric system. There were regions where this conservation was temporarily destroyed and we have noted that such effects
arise due to the skyrmions interaction with the boundary of the system. However for critical values of $\Gamma$ and $d$, during the interaction 
of a skyrmion configuration with a potential hole, there exists regions where ${J}_{tot}$ is not conserved. Here it shall be seen
that the regions where $\dot{J}_{tot} \ne 0$ do not arise due to to boundary effects but are due to an important physical process.

\begin{figure}[!h]

\unitlength1cm \hfil
\begin{picture}(8,8)

\epsfxsize=10cm \put(-1,0.25){\epsffile{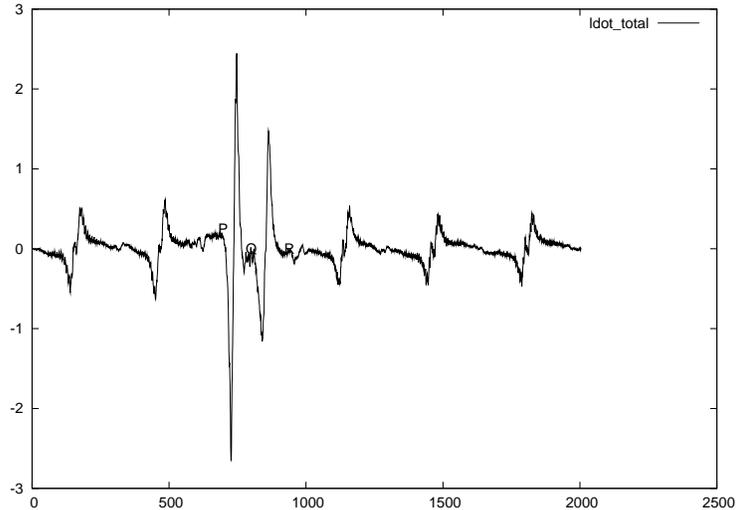}}

\end{picture}
\caption{\label{ldot trans 0.5 total} Plot of the derivative of the total angular momentum, $\dot{J}_{tot}=\dot{l}_{fields}+\dot{l}_{obs}$ 
for a $d=5.5$ two-skyrmion configuration interacting with a potential hole of 
width $b=2$ and  $\Gamma=-0.5$. }

\end{figure}

\begin{figure}[htbp]

\unitlength1cm \hfil
\begin{picture}(10,10)
\epsfxsize=14cm \put(-1,0.25){\epsffile{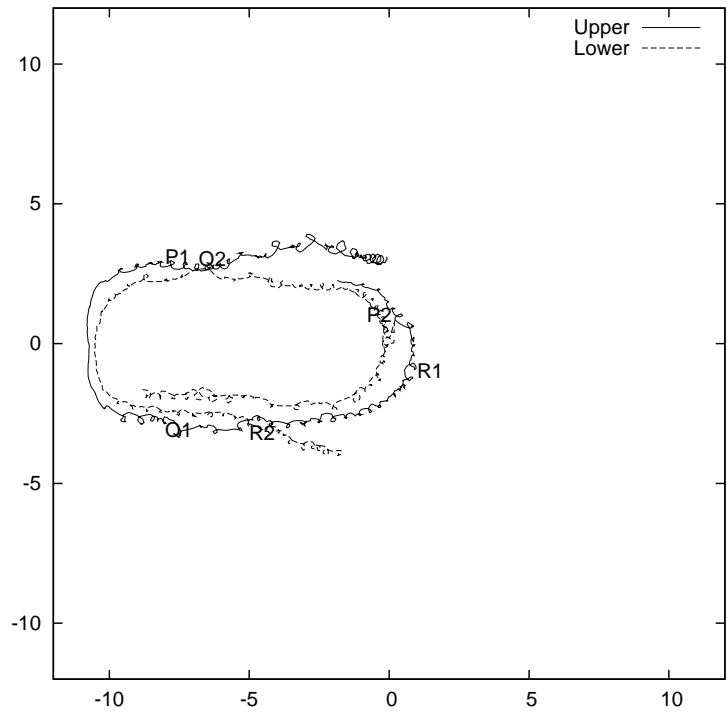}}
\end{picture}
\caption{\label{traj label 0.5}Trajectory of both skyrmions for a $d=5.5$ configuration interacting with a 
potential hole of width $b=2$ and depth $\Gamma=-0.5$, plotted over the time range $500-1000$, where the horizontal 
axis represents the $x$ coordinate and the vertical axis the $y$ coordinate. }

\end{figure}

\begin{figure}[htbp]

\unitlength1cm \hfil
\begin{picture}(14,14)

\epsfxsize=10cm \put(0,8.25){\epsffile{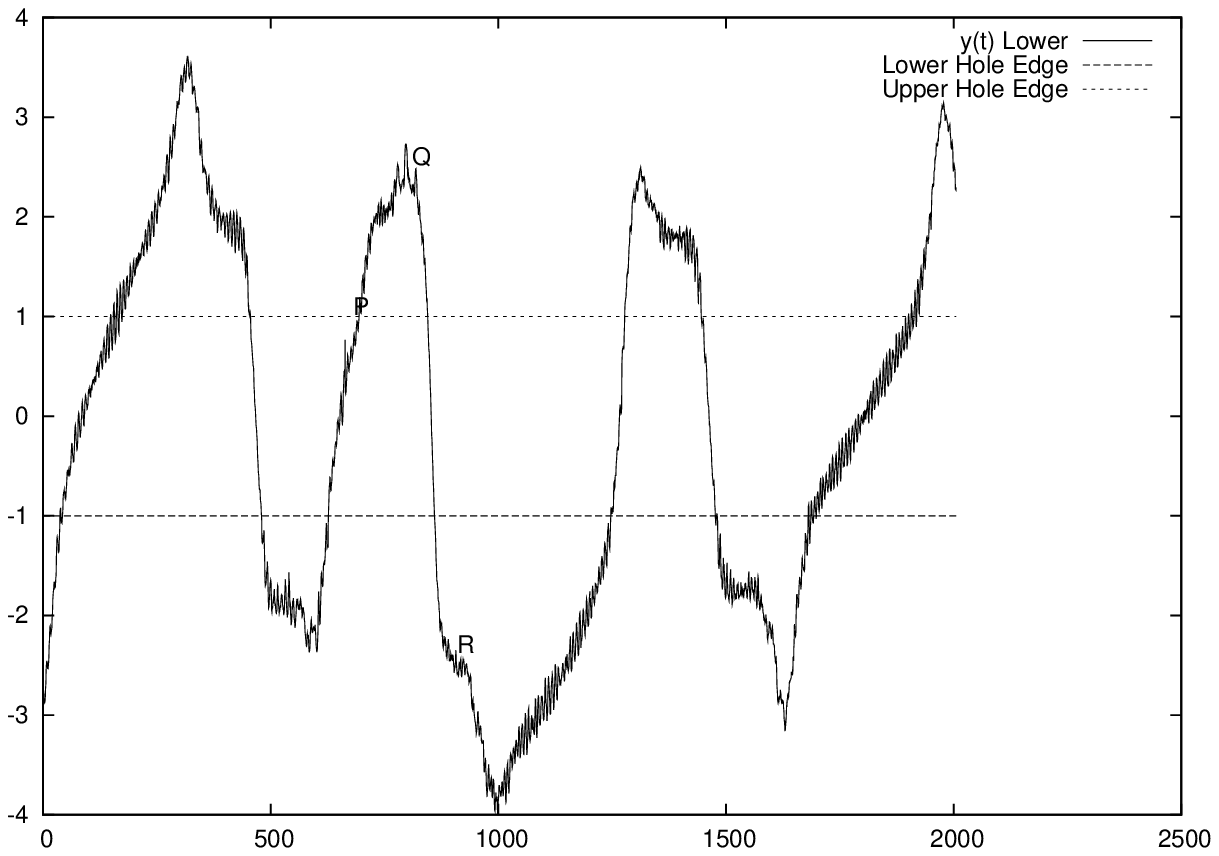}}
\epsfxsize=10cm \put(0,0.25){\epsffile{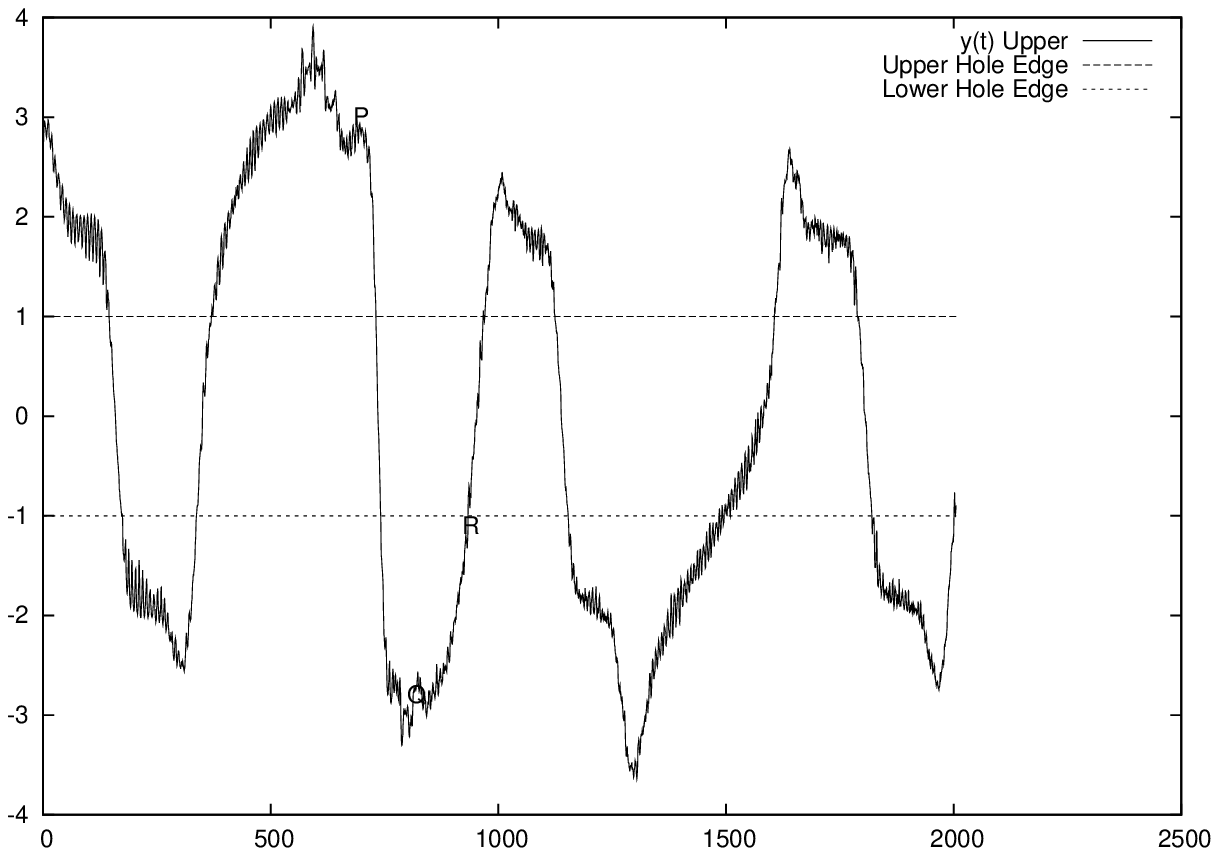}}

\put(5,7.5){a}
\put(5,0){b}

\end{picture}
\caption{\label{upper y(t) 0.5} Plots of $y(t)$ for the upper (b) and lower (a) skyrmions in a system with an asymmetric potential hole of width $b=2$ and  $\Gamma = -0.5$. }
\end{figure}

\begin{figure}[p]
\unitlength1cm \hfil

\begin{picture}(12,12)
 \epsfxsize=6cm \put(0,8){\epsffile{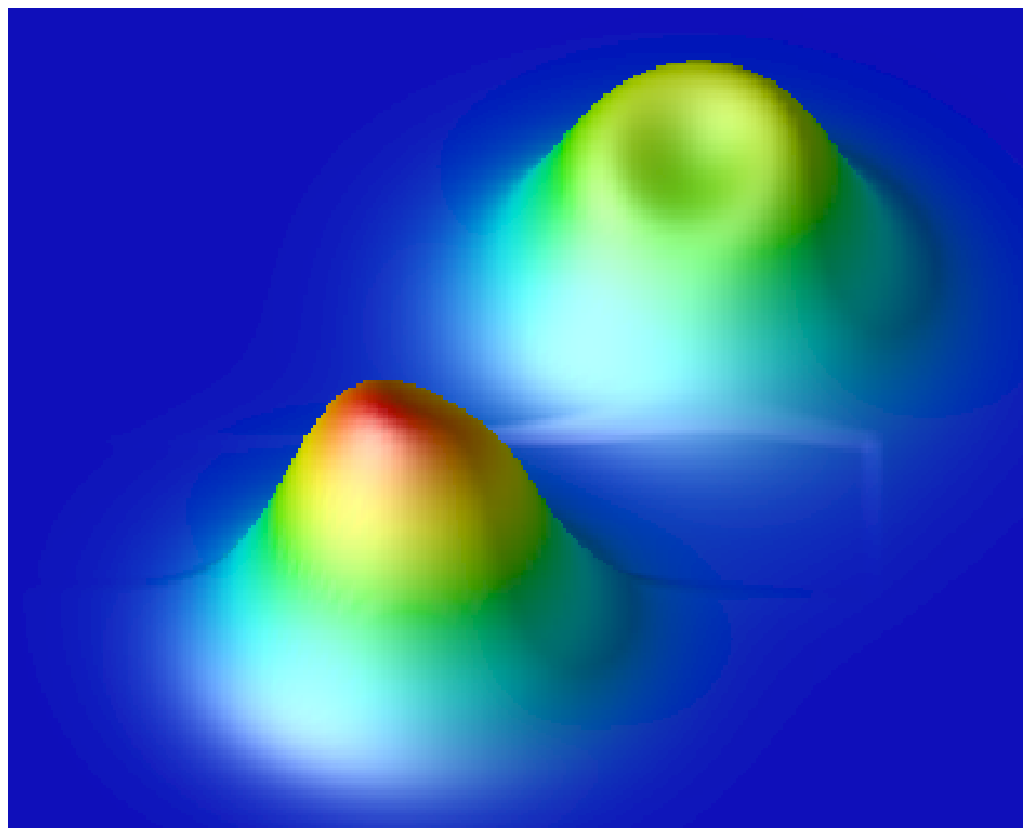}}
 \epsfxsize=6cm \put(7,8){\epsffile{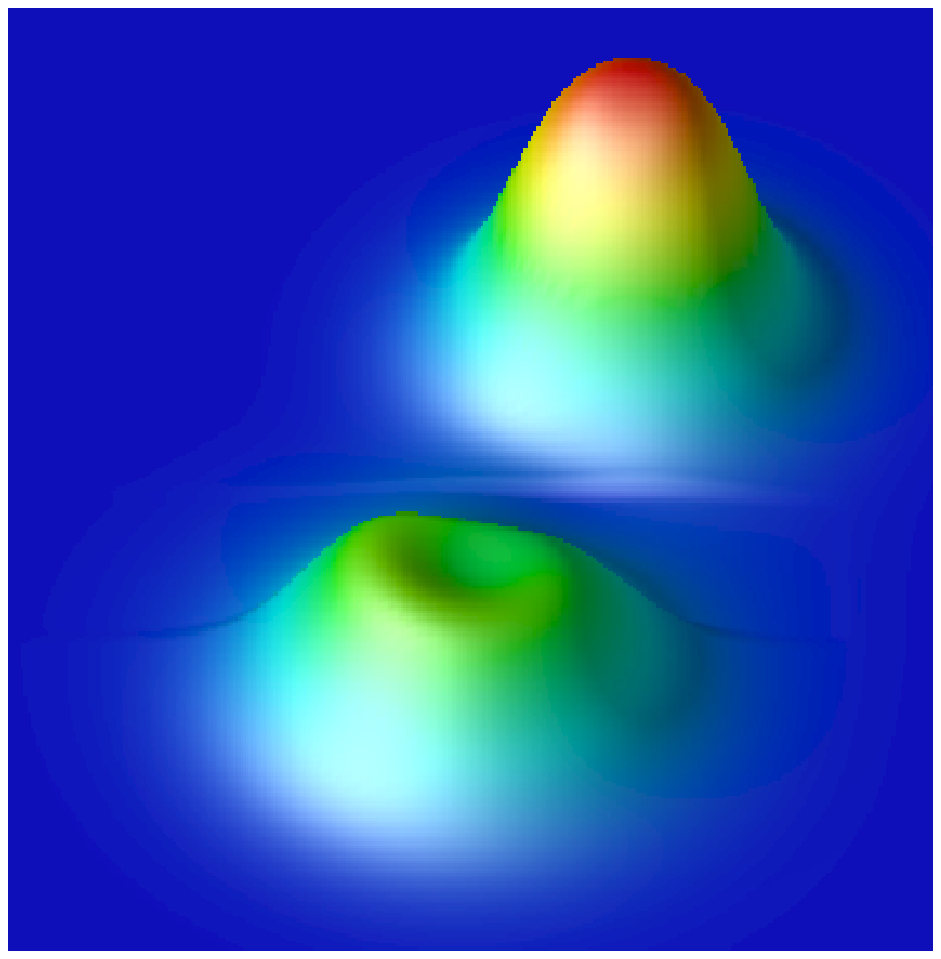}}
 \epsfxsize=6cm \put(0,0.25){\epsffile{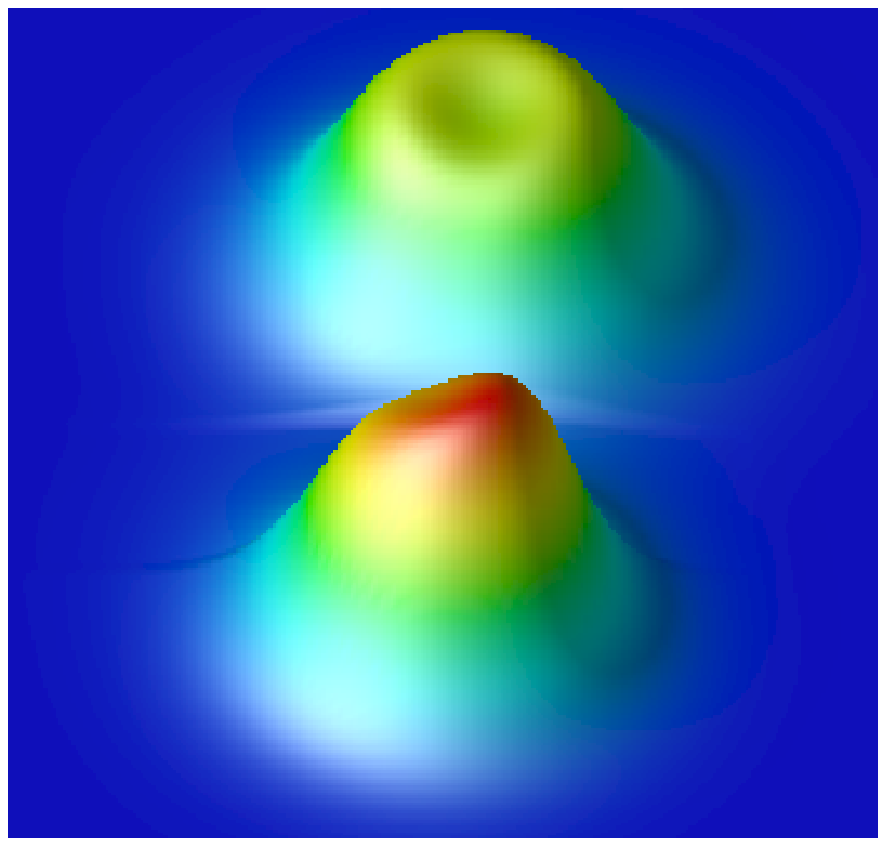}}
\epsfxsize=6cm \put(7,0.25){\epsffile{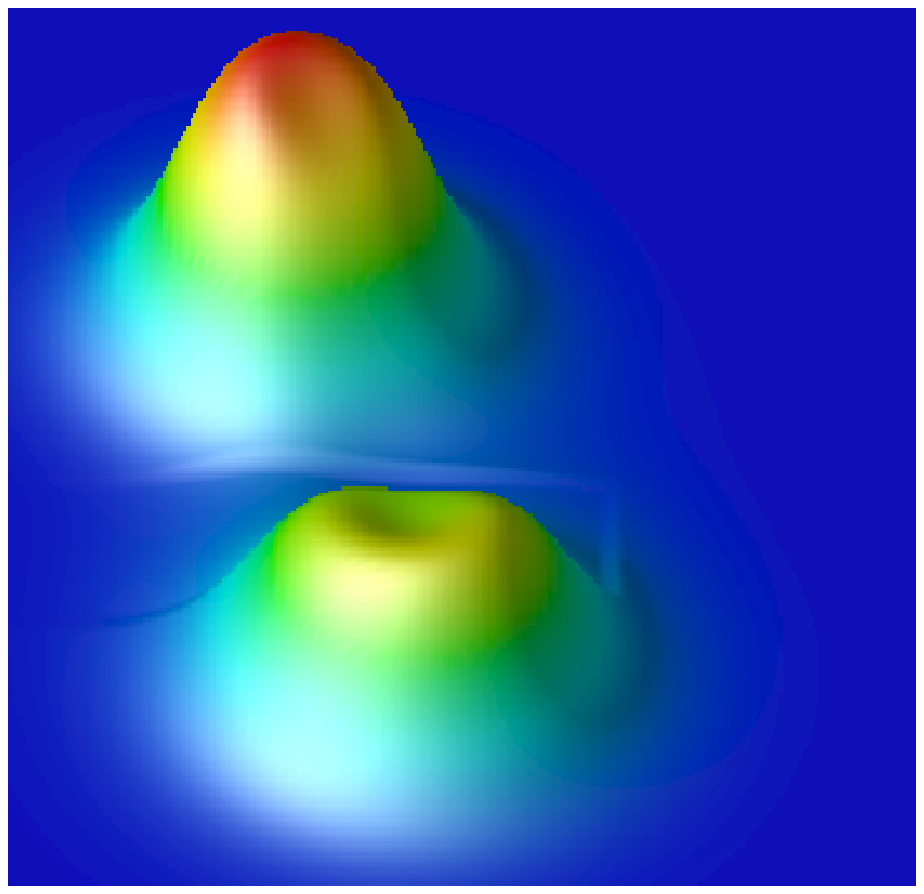}}
\put(3,-0.25){c}
\put(10,-0.25){d}
\put(3,7.5){a}
\put(10,7.5){b}

\end{picture}
\caption{\label{skyrme osc} Plots of the energy densities of the upper and 
lower skyrmions for a $d=5.5$ skyrmion configuration interacting with a potential hole of width $b=2$ and depth $\Gamma=-0.5$. The time line for this set of plots begins with a) through to d) in 10 second increments.}

\end{figure}

  Fig. \ref{ldot trans 0.5 total} shows the plot of $\dot{J}_{tot}$ for a $d=5.5$ skyrmion configuration interacting with potential hole of width $b=2$ and depth $\Gamma=-0.5$.
 The trajectories of both the upper and lower skyrmions in this system are plotted in Fig. \ref{traj label 0.5} over the period of time $t=500-1000$ secs.
 This period corresponds to the same period in Fig. \ref{ldot trans 0.5 total}, when the 
large spike in $\dot{J}_{tot}$ is observed.
 The points P,Q and R of Fig. \ref{ldot trans 0.5 total} can be found by examining Fig. \ref{upper y(t) 0.5},
which show plots of $y(t)$ for both the upper 
and lower skyrmions. The locations of the upper and lower skyrmions for the points P,Q and R are shown on the trajectory 
in Fig. \ref{traj label 0.5} and are given the subscript 1 and 2 to identify the upper and lower skyrmions respectively. During the periods of P-Q and Q-R, 
$\dot{J}_{tot}$ undergoes a rapid change. Over the same time length both the skyrmions have traversed the hole, firstly the upper 
skyrmion, $s_{1}$, and then the lower, $s_{2}$. In the path, P1-Q1, $s_{1}$ traverses the hole while $s_{2}$ starts to move along the edge of the hole. 
In the next path, Q1-R1, $s_{1}$ now moves along the lower edge of the hole and begins to execute circular motion after its 
interaction with $s_{2}$. The path Q2-R2 coincides with $s_{2}$ traversing the hole. 
We must therefore ask ourselves, what distinguishing features are present during the time periods P-Q and Q-R that are absent the rest of the time? To do this 
we examined  a simulation of the energy densities of the skyrmion configuration for the $d=5.5, \Gamma=-0.5$ system, during this time period. Fig. \ref{skyrme osc}
shows snap shots of the energy densities of the upper and lowers skyrmion in this system. The snap shots evolve from a) to d) in 10 second increments.   
In such a system one can see the skyrmions undergo large oscillations in their energy density during their motion. 
In the simulation of this system, $s_{1}$ traverses the hole during the initial $500$ secs. 
After $s_{1}$ traverses the hole for a second time, the skyrmions undergo a violent oscillation in their energy 
densities. Their energy densities dramatically pulse before the skyrmions continue with their motion around the edge of the hole. The first of these 
oscillations occur during the 
period P-Q and the skyrmions are subsequently seen to undergo another oscillation, during the period Q-R. The large change in the energy 
densities, causes the observed spike in Fig. \ref{ldot trans 0.5 total}. There exist other examples of skyrmions oscillating in these systems and in all cases
the conservation of $J_{tot}$ is destroyed during such periods. This evidence leads us to the observation that the 
non-conservation of $J_{tot}$ and skyrmion oscillations in Landau-Lifshitz models are intertwined.

 
\section{Conclusion}

In the asymmetric barrier system the skyrmions behave similarly to the skyrmions observed in the symmetric barrier system. The bound 
skyrmions perform the normal circular motion about the configuration centre moving on and off the barrier during their evolution. The 
distance of separation of the skyrmions increases and decreases with time. The maximum separation between the two skyrmions corresponds
 to the times when one or both of the skyrmions is off the obstruction. The minimum separation corresponds to times when both the skyrmions are 
interacting appreciably with the barrier. This change in the distance of separation between the skyrmions results in the distortion of 
 their path as they traverse the obstruction.  In this same system the skyrmions are able to move into the upper and lower planes. In the
 symmetric system they could not do this due to the obstruction existing over the whole $x$ range. When the skyrmions have moved away 
from the obstruction they orbit the boundary of the system. It was noted that this phase of the dynamics is uniquely determined by the
 boundary of the system. If the skyrmions are placed in an identical system, with a larger grid, after initially executing identical 
trajectories as the skyrmions for the system of a smaller grid, the skyrmions move in a near identical motion about the boundary of
 the system, except they are further from the obstruction than the skyrmions of a smaller grid. The binding energy in the barrier 
system also plays an important role. The bound state formed by two skyrmions can be broken by a potential barrier. The breaking point
 in the asymmetric system is much more difficult to observe than in the symmetric system. In the symmetric system the geometry of the potential 
obstructions uniquely determines the trajectories of the unbound skyrmions. The unbound skyrmions in the asymmetric system move 
almost independently of each other about the system. Periodically the 
skyrmions, when they are close enough to interact due to the confinement of the boundary, execute circular motion about their centre. 

In the system of a potential hole the skyrmions behave completely differently from the skyrmions observed in the symmetric system. 
Here the binding energy plays a crucial role in categorising the different phases of the dynamics; bound, unbound or in between the 
two. The skyrmions, for tightly bound configurations, orbit each other as they traverse the edge of the hole. These bound states 
occur for small distances of separation and hole depth. As the hole depth or the distance of separation increases, the binding 
energy of the skyrmions increases. However, as the skyrmions approach the point at which the bound state is broken, the dynamics 
of the skyrmion configuration changes. The skyrmions in these `in between states' are able to penetrate the hole. They behave as
 bound skyrmions in the sense that their trajectory could be described as a very distorted circular path but they behave 
as unbound skyrmions since they separate
 appreciably during the simulation. In a potential hole system, as was noted in the analysis on the symmetric system, the binding
 energy becomes a dynamic quantity. As the skyrmions approach the hole the binding energy or the interaction energy between the 
skyrmions, changes.
The skyrmions are extended objects. When one of the skyrmions approaches the hole, the energy of the skyrmion 
decreases due to the tail interactions with the hole. Due to the asymmetry of the system, one of the skyrmions can feel the effect
 of the hole more than the
other. Energy conservation allows the skyrmions to separate. The more a skyrmion approaches the hole the more they can separate
 and thus the skyrmion bound state is broken by the potential hole. This process is sensitive to the hole depth, $\Gamma$. In the
 transition state dynamics, $d=5.5$, increasing the value of $\Gamma$ results in a complete change in the behaviour of the 
skyrmions. This is largely due to the binding energy as the skyrmion initially approaches the hole. In the case of large values 
of $\Gamma$, the skyrmions know they are unbound when they are further from the hole and begin to execute dynamics reminiscent 
of the skyrmions for a symmetric potential hole. When $\Gamma$ is slightly smaller, due to $E_{B}$, the skyrmion is able to penetrate  
 the hole and there is still an appreciable interaction with the other skyrmion. This allows the skyrmion
 to escape the hole and a pseudo-bound state behaviour to resume. Our examination of the dynamic behaviour of the binding energy
 led us to a reasonably accurate estimate for the closest initial distance of approach of the upper skyrmion during the transition
 dynamics. We thus believe that the subsequent analysis regarding the binding energy and the skyrmions' dynamics dependence upon
 it, is a valid and a good representation of the details of these complicated systems. 

The system of unbound skyrmions in the potential hole system also shows some unusual properties, some of which were seen in the
 transition dynamics. The unbound skyrmions occur for large values of the distance of skyrmion separation and large hole depths. 
The skyrmions of these system behave, as was alluded to previously, similarly to the skyrmions in the symmetric potential hole 
system where both skyrmions moved along the axis of the hole. Here, due to geometry of the hole, only one of the skyrmions moves 
along the edge of the hole, while the other, after the initial motion, comes to rest. During the initial period of separation 
both of the skyrmions' energy densities are seen to oscillate. This oscillation continues until the skyrmions reach a critical 
distance of separation, where it ceases. These oscillations surface again after the skyrmion is reflected from the boundary, 
when it has reached another critical distance from the stationary skyrmion. These oscillations are rather peculiar and the only 
other system to bear such oscillations is the transition dynamics. In the transition dynamics for $\Gamma= -0.4$ and $ -0.5$, 
the skyrmions' energy densities undergo violent oscillations. These oscillations are the result of the excitation of one or more
 of the internal modes of the skyrmions due to the energetics during the skyrmion separation. We would expect that any of the 
excited modes would be of the lowest order. Piette et al \cite{PW05} have previously undertaken an analysis of the excited modes of 
the skyrmions of the new baby skyrme model, in the fully relativistic system. In \cite{PW05} the modes of a single skyrmion and
 a two-skyrmion configuration during a scattering process were analysed. It would be interesting to see how the excited modes of the skyrmions
 observed in our simulations compare with the excited modes seen in their work.

One of the most interesting observations of our simulations was the apparent non-conservation of the total angular momentum $J$ 
(given its usual definition). This non-conservation of $J$ was due to the non-conservation of the orbital angular momentum, $l$, 
as we have found that in all of the simulations the total magnetization in the third direction, $m$, was well conserved in time. 
At the same time we showed  that $\dot{l} \ne 0$.
Thinking about this further we showed that when a system involves potential obstructions, these obstructions made a significant 
contribution to $\dot{l}$. Hence one has to modify the conventional definition of $l$. We have found and calculated this missing 
contribution. In all of the simulations of the skyrmions in both the symmetric and asymmetric systems we have shown that the addition 
of the contribution from the obstructions compensates $\dot{l}$. This results in the overall conservation of $l$ and $J$ for the full system. 
However, there were a few systems
 studied that didn't adhere to the above statement. In the transition dynamics in the asymmetric system, there were regions where the
 angular momentum due to the fields, $\dot{l}_{fields}$, and the angular momentum due to the obstruction,
$\dot{l}_{obs}$, did not sufficiently cancel each other out, in order for total angular momentum conservation. However, when 
comparing the size for which $\dot{J} \ne 0$ in such regions with the total angular momentum as a whole, 
these regions of non-conservation are brief and the magnitude in most of the cases is small.
 In some systems, where the non-conservation of $J$ was larger,
we noted that this was due to boundary effects.

 An examination of the times when there was an appreciable discrepancy in $\dot{l}_{fields}+\dot{l}_{obs} \ne 0$,
 provided a direct link between the skyrmion oscillations and the 
non-conservation of $J$. Initially $J$ is well conserved but as the oscillations begin to increase this conservation is destroyed. 
There reaches a point when the $J$ conservation is restored. It is at these points that the skyrmion oscillations have ceased. 
In many of the cases these oscillations reappear again and again during the skyrmions dynamics and each time the conservation of
 $J$ is destroyed. The size of the oscillation and the magnitude to which $\dot{J} \ne 0$ are also correlated. In the transition 
dynamics the large spike observed in the $\Gamma= -0.5$ plot of $\dot{l}_{fields}+\dot{l}_{obs}$  system corresponds to the violent
 oscillation during the dynamics of the skyrmions. Conversely in the system with $\Gamma= -0.3$ smaller oscillation are apparent
 in that system, resulting in a smaller spike in  $\dot{l}_{fields}+\dot{l}_{obs}$.  The regions where $\dot{J}_{tot} \ne 0$ 
suggest that the inclusion of an additional term is
 required for the overall exact conservation of the total angular momentum, $J$, for skyrmion scattering in a Landau-Lifshitz model.
Any term or terms that are included must
 have a direct correspondence with the oscillations of the skyrmions observed in these systems. 

\section{Appendix: Numerical procedures}
Unfortunately, it is impossible to solve (\ref{LL}) analytically we have therefore had to study this problem numerically. The fields and their derivatives were discretised in the usual manner and were placed on a lattice of $251\times251$ points, with lattice spacing $dx=0.1$. The numerical integration of the 3-coupled differential equations of (\ref{LL}) involved the use of a $4^{th}$ order Runge-Kutta method 
of simulating time evolution with a time step of $dt=0.001$. The various integrals calculated throughout the simulations were performed using a 2-D Simpson's rule. The constraint equation requires that the fields lie on the 2-sphere, $\underline \phi^{2}=1$, and this was imposed at every time step by rescaling each field component so that $\phi_{i}\rightarrow\frac{\phi_{i}}{\sqrt{\underline \phi \cdot \underline \phi}}$.
 
The skyrmions were initially placed at $(0,\pm d/2)$ in the upper and lower planes, where d is the distance between the two skyrmion centres $(x_{i},y_{i})$. The trajectory of each skyrmion was tracked by following the maxima of the topological charge density and interpolating between the lattice points. The coefficients $\gamma_{i}$ have been set to unity in all the simulations unless stated otherwise.

\end{document}